\documentclass[onecolumn]{aastex61}

\usepackage{amsmath}
\usepackage{pifont}
\usepackage{selectp}
\usepackage{mathtools,amssymb}
\usepackage{multirow}
\usepackage[utf8]{inputenc}
\usepackage{nicefrac}
\usepackage{xfrac}
\usepackage{graphicx}
\usepackage{comment}

\usepackage[T1]{fontenc}

\newcommand\be{\begin{equation}}
\newcommand\ee{\end{equation}}

\def\kms{\ifmmode{\rm km\thinspace s^{-1}}\else km\thinspace s$^{-1}$\fi}

\begin{document}

\pagenumbering{arabic}

\shorttitle{Haumea Family Formation}
\shortauthors{Proudfoot \& Ragozzine}

\title{Modeling the Formation of the Family of the Dwarf Planet Haumea}

\author[0000-0002-1788-870X]{Benjamin C. N. Proudfoot}
\author[0000-0003-1080-9770]{Darin Ragozzine}
\affiliation{Brigham Young University, Department of Physics and Astronomy, N283 ESC, Provo, UT 84602, USA}

\email{benp175@gmail.com, darin\_ragozzine@byu.edu}

\setcounter{footnote}{0}

\begin{abstract} 

The dwarf planet (136108) Haumea has an intriguing combination of unique physical properties: near-breakup spin, two regular satellites, and an unexpectedly compact family. While these properties indicate formation by collision, there is no self-consistent and reasonably probable formation hypothesis that can connect the unusually rapid spin and the low relative velocities of Haumea family members ("Haumeans"). We explore and test the proposed formation hypotheses (catastrophic collision, graze-and-merge, and satellite collision). We flexibly parameterize the properties of the collision (e.g., the collision location) and use simple models for the three-dimensional velocity ejection field expected from each model to generate simulated families. These are compared to observed Kuiper Belt Objects using Bayesian parameter inference, including a mixture model that allows for interlopers from the background population. After testing our methodology, we find the best match to the observed Haumeans is an isotropic ejection field with a typical velocity of 150 m s$^{-1}$. The graze-and-merge and satellite collision hypotheses are disfavored. Including these constraints, we discuss the formation hypotheses in detail, including variations, some of which are tested. Some new hypotheses are proposed (a cratering collision and a collision where Haumea's upper layers are "missing") and scrutinized. We do not identify a satisfactory formation hypothesis, but we do propose several avenues of additional investigation. In addition, we identify many new candidate Haumeans and dynamically confirm 7 of them as consistent with the observed family. We confirm that Haumeans have a shallow size distribution and discuss implications for the identification of new Haumeans. 

\end{abstract}


\section{Introduction}
\setcounter{footnote}{0}

The core accretion hypothesis of solar system formation implies that collisions between icy bodies were extremely common. As an integral part of planet formation in the outer solar system, improving our understanding of icy collisions requires advances in numerical modeling, laboratory analyses, and observational constraints. In the asteroid belt, the study of collisional families has developed hand-in-hand with detailed numerical simulations, each informing the other. Less is understood about the physics of icy collisions, and empirical constraints from observational analyses of families of Kuiper belt objects (KBOs) are essential. 

At present, the only KBO family is associated with the dwarf planet (136108) Haumea (\citealt{2007Nature..446..296}, \citealt{2011ApJ...733...40M}). Besides a concentration of similar orbits that show unique deep water ice features (\citealt{2007Nature..446..296}, \citealt{2008ApJ...684L.107S}, \citealt{snodgrass2010characterisation}, \citealt{carry2012characterisation}), Haumea is also the fastest spinning large\footnote{Using the compilation of rotational periods from the Light Curve DataBase (LCDB; \citealt{2009Icar..202..134W}) queried on October 5, 2018, the largest body with a well-known rotational period shorter than Haumea's is (201) Penelope, which is an order of magnitude smaller. There are some large (few hundred km) KBOs with likely rotational periods comparable to Haumea's 3.915 hours.} body in the solar system with two moons that are on orbits that are too coplanar to form through capture \citep{2009AJ....137.4766R,ortiz2017size}. The combination of these unique properties suggests that the rapid spin, satellites, and family of Haumea formed in a single collision, but identifying a single self-consistent hypothesis has been challenging (e.g., \citealt{bagatin2016genesis}). 

A primary constraint on the formation of the Haumea family is the distribution of Haumea family members, hereafter "Haumeans". A major challenge in explaining the Haumea family is the small velocity kick ($\boldsymbol{\Delta v} = (\Delta v_x, \Delta v_y, \Delta v_z$)) for spectrally-confirmed Haumeans, $\Delta v \lesssim$150 m s$^{-1}$, an order of magnitude smaller than the velocities expected in a catastrophic collision  -- $\sim$3 times the escape velocity, i.e. $\simeq$ 2700 m s$^{-1}$. Furthermore, collision modeling by Smoothed Particle Hydrodynamics (SPH) shows that near-breakup rotation is not consistent with a large catastrophic collision (e.g., \citealt{2010ApJ...714.1789L}, \citealt{2012MNRAS.419.2315O}). This well-known result has led several authors to propose alternate formation hypotheses for the Haumea family. These hypotheses can roughly reproduce the low velocities, though the plausibility of these hypotheses requires further scrutiny. 

In this work, we seek to explore and test the proposed formation hypotheses in detail. After discussing Haumea formation hypotheses (Section \ref{sec:hypotheses}), we explain how we generate simulated families using a flexible parametrization and the unique aspects of each model (Section \ref{sec:models}). We identify new candidate Haumeans (Section \ref{sec:observed}) and explain how we use Bayesian parameter inference to match the distribution of observed objects with our simulated models (Section \ref{sec:bayesianfit}). After establishing the validity of our models (Section \ref{sec:testing}), we describe our Results for each model in Section \ref{sec:results}. We use our preferred model to confirm new Haumeans (Section \ref{sec:newhaumeans}) and discuss implications for the number and size distribution of the family (Section \ref{sec:numbersize}). Each of the formation models is discussed in detail (Section \ref{sec:discussion}) and other formation hypotheses are considered (Section \ref{sec:otherhypotheses}). Conclusions and future work are presented in Section \ref{sec:conclusions}. 


\subsection{Haumea Formation Hypotheses}

\begin{figure}
    \centering
    \includegraphics[width=0.9\textwidth,keepaspectratio,trim={0 4in 0 0} ]{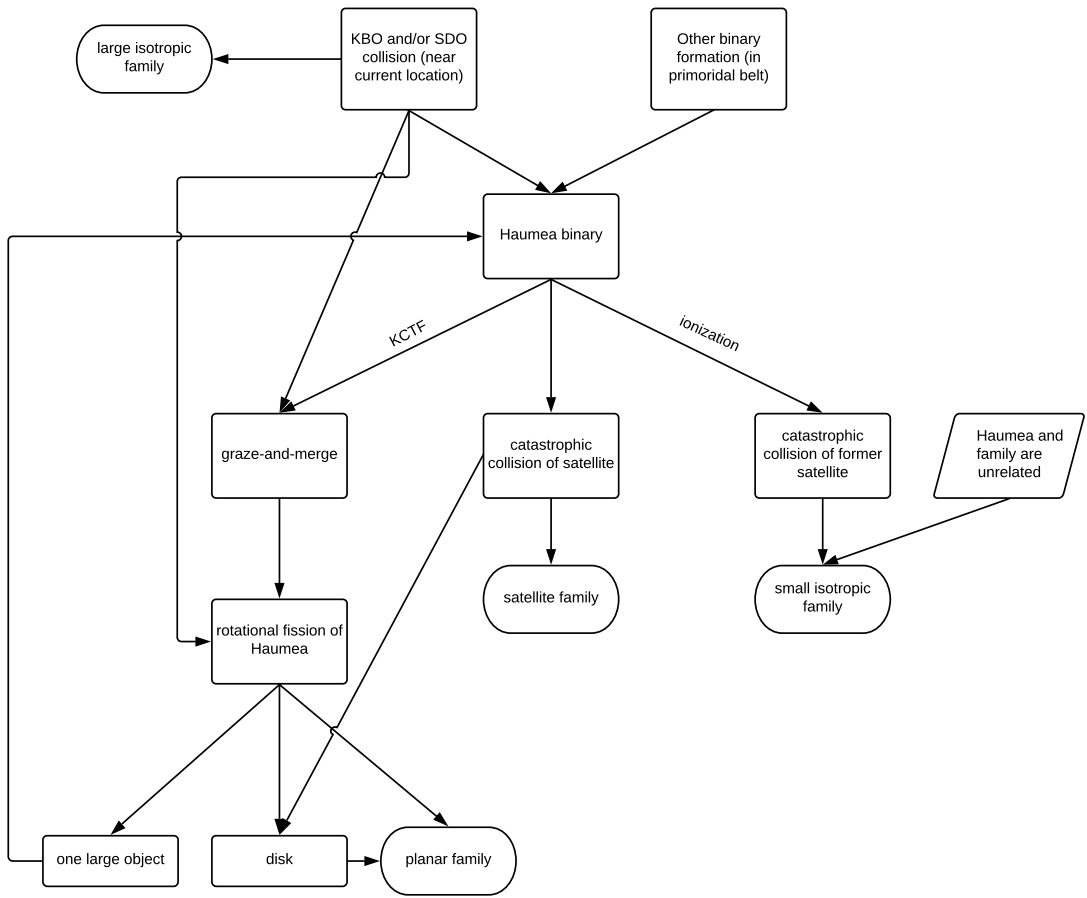}
    \caption{A graphical summary of the main hypotheses proposed for the formation of Haumea's family. Various authors have proposed different pathways through this diagram considering the probability of each step based on known or estimated physical mechanisms. For example, we elaborate on the formation of a planar Haumea family by starting with the formation of a near-equal binary in the primordial belt that, upon placement in Haumea's current high-inclination orbit, undergoes Kozai Cycles with Tidal Friction (KCTF) that leads to a reasonably probable graze-and-merge type formation, leading to the rotational fission of Haumea which forms a planar family. We identify four major outcomes for the final shape of the Haumea family: a large (e.g., comparable to Haumea's escape velocity of $\sim$1 km s$^{-1}$) isotropic family, a small ($\sim$150 m s$^{-1}$) isotropic family, a satellite family (sculpted by interactions with Haumea), and a planar family where excess angular momentum leads to ejection of objects primarily in plane. Each of these models has a different shape in the $\mathbf{\Delta v}$ distribution.}
    \label{fig:hypotheses}
\end{figure}

\label{sec:hypotheses}
Since the discovery of the Haumea family, several formation hypotheses have been proposed. For the most part, we focus on formation hypotheses that attempt to explain the spin, satellites, and family in a single event or a probable sequence of events.\footnote{The formation and evolution of Haumea's rings are not well enough understood to use them as a constraint; for example, their mass is not known. The rings may not be long-lived. Perhaps recent small collisions or even endogenic changes are sufficient to produce the rings. We thus do not include ring formation as a constraint on family formation hypotheses.} Invoking completely separate formation mechanisms for a rare spin rate and a rare collisional family significantly reduces the probability of matching the observations (e.g., \citealt{bagatin2016genesis}, \citealt{2016AJ....152..195H}). In other words, we restrict our focus on hypotheses that connect the dynamically tight group of objects with unique deep water ice spectra (e.g., Haumeans) with the dwarf planet Haumea, which has a similar spectrum, was plausibly located near the center of this collision (before diffusion in its current 12:7 resonance, \citealt{2007AJ....134.2160R}), and which shows signs of a unique collision (near rapid breakup and two distant satellites). 

The multiplicity of hypotheses stems from the goal of identifying a model that matches all the observed constraints, is reasonably probable, and shown to be physically plausible through simulations (or at least rough estimates). Many interrelated variations are now available in the literature -- as illustrated in Figure \ref{fig:hypotheses} -- such that a simple categorization scheme is not possible. As our primary goal in this paper is to provide new constraints based on the shape of the family, we introduce the expected shapes through a brief discussion of the more common proposed hypotheses. 

\subsubsection{Catastrophic collision of Haumea} 
In the Haumea family discovery paper, \citet{2007Nature..446..296} suggest that Haumeans might result from a "standard" catastrophic collision between two large bodies\footnote{Here, and throughout this work, we refer to catastrophic collisions. By this we mean the shattering of the target as is common in asteroid collisions, rather than a cratering event.}, though with the recognition that this would not match the compact $\Delta v$ distribution of proposed family members. \citet{2008AJ....136.1079L} pointed out that the required collision between two large bodies in the current\footnote{The apparently pristine distribution of Haumeans cannot be reasonably preserved if the family forms elsewhere and is "moved" to the present location. Because any process that moves objects would not be 100\% efficient, the tight cluster of all objects with strong water ice spectra is inconsistent with significant transport.} Kuiper belt was very improbable due to the low number densities in the dynamically hot population. They proposed that the collision of two scattered/scattering disk objects could result in the formation of the observed family, allowing them to invoke the fact that this population used to be $\sim$100 times more populous, rendering such a large collision reasonably probable. 

These models cannot reasonably account for the low values of $\Delta v$ seen for Haumea. Catastrophic collisions as a rule create ejection velocities comparable to the escape velocity of the target \citep[e.g.,][]{2012ApJ...745...79L} with velocities of $\sim$3 times higher than the escape velocity being typical. With velocities of $\sim$0.1 times the proto-Haumea's escape velocity, the known Haumeans cannot be explained in any standard collision model. A first principles argument also supports this conclusion: for Haumeans to end up with such low $\Delta v$ implies that large numbers of objects had their initial (near-Haumea) velocities in a range of $\sim$1.0-1.1 times the escape velocity, requiring incredible fine tuning. 

Simulations of catastrophic collisions show that they result in an essentially isotropic distribution of fragments in the center of mass frame. Hence, such collisions form a large, isotropic family. 

\subsubsection{Satellite destruction hypothesis}

To explain the low velocity dispersion of Haumeans, \citet{2009ApJ...700.1242S} propose that the Haumea family formed in the destruction of a former satellite of Haumea (called the "ur-satellite" by \citealt{cuk2013dynamics}). A smaller satellite would have a lower escape velocity and the $\Delta v$ values are not dissimilar to the orbital velocities of the known satellites \citep[70-90 m s$^{-1}$,][]{2009ApJ...698.1778R}. 
\citet{cuk2013dynamics} also prefer the satellite collision model as the first step for forming Haumea's satellites. 

\citet{2009ApJ...700.1242S} propose that the ur-satellite was destroyed by a heliocentric KBO impactor and show that such a collision is plausible with current number densities over the age of the solar system. This would create a unique shape in the ejection velocities of the Haumea family since the destruction of a satellite while in orbit around Haumea leads to a velocity boost and interactions of the objects with Haumea (discussed further below). 

\citet{2012MNRAS.419.2315O} and \citet{bagatin2016genesis} propose variations on this hypothesis, where the ur-satellite is formed through shedding of excess angular momentum induced by a collision (discussed in more detail below). 

They also point out that Haumea's ur-satellite may have been already ionized from Haumea and was then destroyed. (Or that fission produces an fragment that never orbits Haumea and is thus never was a "satellite".) This maintains the connection between the family and Haumea (though would require renaming as it is no longer the "Haumea" family), while producing a small isotropic family. \citet{bagatin2016genesis} also investigate whether Haumea and the "Haumeans" must be related at all, suggesting that destruction of a smaller object could be coincidental with Haumea's proper orbital elements.

The primary disadvantage of the satellite collision hypothesis is that it usually decouples the formation of Haumea's unique spin and Haumea's unique family \citep[e.g.,][]{bagatin2016genesis}. 
This model does not require that Haumea's spin be near-break up and, in fact, the formation of the large ur-satellite is generally inconsistent with rapid rotation in collisional modeling \citep{2010ApJ...714.1789L,2012MNRAS.419.2315O}, discussed in greater detail below. 

The mass of the Haumeans is now known to be $\sim$3\% of the mass of Haumea (\citealt{vilenius2018tnos}, Pike et al. 2019, submitted). While previous formation hypotheses were content to say that the ejection velocity from a catastrophic collision is the same or very similar to the escape velocity of the surface, we require the more accurate result that the ejection velocity is typically a few times the escape velocity with a wide distribution \citep{2012ApJ...745...79L}. As discussed in more detail below, the mass of the Haumeans is much too large to be self-consistent with such a catastrophic collision. This is a serious drawback to any model that attempts to explain the tight distribution of Haumeans by a typical catastrophic collision.

\subsubsection{Graze-and-Merge Formation Hypothesis}

\citet{2010ApJ...714.1789L} propose that a low velocity grazing collision between two large objects would cause these objects to merge into a single object with so much spin angular momentum that it would shed mass off the tips of the triaxial body at low velocities, thus forming the Haumea family. This model naturally connects Haumea's near-breakup spin with the small family and is shown to be physically self-consistent in SPH simulations by \citet{2010ApJ...714.1789L}. 
A critical flaw of the graze-and-merge hypothesis is that it severely exacerbates the issue of having such a collision in the current Kuiper belt because it requires an unreasonably low impact velocity between two very large objects. However, \citet{2011ApJ...733...40M} posit that the proto-Haumea could have started as a near equal mass binary that forms in the primordial disk. Kozai Cycles with Tidal Friction (KCTF) \citep{porter2012kctf} could naturally destabilize the proto-Haumea binary once it is emplaced (by Neptune scattering) in its high inclination orbit, allowing for a family formation that occurs separately the binary formation. The impact velocities of a bound binary are similar to the near-zero relative velocity collisions simulated by \citet{2010ApJ...714.1789L}. This proto-binary variation on the graze-and-merge hypothesis renders the collision plausible with reasonable probability of occurring in the actual Kuiper belt. 
    
\citet{2012MNRAS.419.2315O} suggest a variation on this model where a series of small (and thus much more probable) collisions could still produce the relevant outcome of an icy body with too much angular momentum shedding its icy mantle at low velocities. This requires the proto-Haumea to be rapidly rotating which is plausible. Their preferred collisionally-induced fission results in an ejection field that can have a small (or large) dispersion around a predominant direction. 

Families that form as a result of rotational fission are strongly constrained to a planar ejection; the models of \citet{2010ApJ...714.1789L} indicate an inclination dispersion of $\lesssim$5$^{\circ}$ around a common plane. 

\section{Methods}
\label{sec:methods}
Each of the proposed formation hypotheses makes a prediction for the distribution of the ejection\footnote{Throughout this paper, we use "ejection" velocity and direction to refer to $\boldsymbol{\Delta v}$ which is the "final" difference in heliocentric velocity vectors. We do not calculate actual velocities after the moment of impact or escape velocities, but instead ejection and $\boldsymbol{\Delta v}$ refer to the final "velocity at infinity" of outgoing family members.} \emph{directions} that have never before been compared to the observations. Variations on the formation hypotheses shown in Figure \ref{fig:hypotheses} can lead to different outcomes, but we can still associate the shape of the family with the physical formation mechanism. For example, rotational fission predicts a highly planar ejection distribution. By using the full $\boldsymbol{\Delta v}$ instead of just the typical magnitude, we can provide additional constraints to help us distinguish between these competing hypotheses. 

To provide a basis for comparing these hypotheses to the observations, we use a simplified model of these mechanisms to create simulated collisional families that are then compared to the properties of the observed Haumeans, including several new candidates. 

We explore a very wide variety of possible models by casting this as a parameter inference problem, though in practice we are sometimes more interested in understanding the plausibility of these models than the specific parameter values that are inferred. With the goal of exploring a variety of models, we elect to use a very flexible parametrization of the families that would result from the various formation hypotheses. 

We first discuss parameters that are common to all models and then discuss parameters that apply to specific models. 

\subsection{Parameters Used in All Models}
\label{sec:models}
Though the orbital parameters of known Haumeans are clustered in space, the exact location of the center of the collision is not precisely known, since Haumea has experienced significant diffusion in the 12:7 resonance (\citealt{2007Nature..446..296}, \citealt{2007AJ....134.2160R}). We therefore include as parameters the orbital elements of the collision center: semi-major axis $a_c$, eccentricity $e_c$, inclination $i_c$, argument of periapse $\omega_c$, longitude of ascending node $\Omega_c$, and mean anomaly $M_c$. These set the position and velocity of the center of the collision, effectively the same as the "collision orbit" or "center of mass" orbit, which should be slightly offset from the final position of Haumea after the collision. As pointed out by \citet{2008AJ....136.1079L}, it also does not correspond to the orbit of the proto-Haumea (assuming an external impactor). As is well known in asteroid family modeling and as confirmed by our analyses, $\Omega_c$ does not contribute to the final distribution of observable quantities, so it is held fixed at 0$\degr$. 

We then produce an arbitrary number of simulated family members by assigning them a $\boldsymbol{\Delta v}$ consistent with the proposed model. All the models share the same parameterization of the size-velocity distribution, so we describe that here. Our methods were inspired by and closely follow \citet{lykawka2012dynamical}. 

Each object was first given an absolute magnitude between 3.4 and 6.0 from a power-law distribution with a slope given by $\alpha$, a free parameter in our models. The range of absolute magnitudes is chosen to range from the largest known Haumean, 2002 TX$_{300}$ to the smallest objects likely to have been detected. The differential absolute magnitude distribution slope $\alpha$ is used in the same sense as it is typically used in studies of the Kuiper belt:
\begin{equation}
\label{h_equation}
  dN/dH \propto 10^{\alpha H} 
\end{equation}
where $\alpha$ is related to the size distribution slope $q$ by $q = 5\alpha +1$, assuming that albedos are constant. Our use of a single power law is well justified based on the results of Pike et al. (2018, submitted). 

Asteroid families and SPH simulations of catastrophic collisions indicate that there is a rough equipartition of momentum among escaping fragments, so that there is a correlation between a family member's size and velocity. Following \citet{lykawka2012dynamical} (but not matching their notation), this correlation was incorporated by a defined fiducial velocity magnitude $v_{fid}$ by 
\begin{equation}
    \label{lykawka1}
    v_{fid}(H) = S\times10^{H\beta}
\end{equation}
where $S$ and $\beta$ are free parameters and $H$ is the absolute magnitude of the simulated family member.

Though \citet{2007AJ....134.2160R} and \citet{2011ApJ...733...40M} model KBO families with a single fixed value of $\Delta v$, this is not realistic and leads to unphysical artefacts in the simulated families. As a result, we randomly scale each fiducial velocity using 
\begin{equation}
    \label{lykawka2}
    \Delta v_{init} = v_{fid}(H)p(x)
\end{equation}
where $p(x)$ is a Weibull distribution
 \begin{equation}
    \label{weibull}
    p(x;\lambda,k) = \frac{k}{\lambda}\left(\frac{x}{\lambda}\right)^{k-1}e^{-(x/\lambda)^{k}}
\end{equation}
where both $\lambda$ and $k$ are free parameters in our model, with reasonable values of $\lambda =0-15$ and $k = 0.75-5$. This gives significant flexibility to the shape, center, and width of the velocity ejection distribution, though since the Weibull distribution is positive definite, this places a minimum value of $v_{init}=v_{fid}(H)$. 

Each individual model takes $\Delta v_{init}$ and combines it with other calculations to develop a final $\boldsymbol{\Delta v}$. This is then added to the heliocentric velocity of the collision orbit. Combined with the (unchanged) position of the collision orbit, the orbital elements of each simulated family member can be calculated, especially the semi-major axis $a$, the eccentricity $e$, and the inclination $i$. The orbital angles, especially $\omega$ and $\Omega$, typically have a very tight distribution $\lesssim$10$^{\circ}$.

It is important to note that these simulated orbital elements are technically osculating elements, even though we will be comparing them to proper (time-averaged) elements (see below). In this analysis, the orbital elements of Haumeans at the time of the collision cannot be determined, even though their proper elements are generally well defined. Comparing simulated osculating elements to observed proper elements could seem problematic, but we propose that this is not a serious limitation as long as we limit ourselves to non-resonant stable objects. 

In Lagrange-Laplace secular dynamics (which forms the basis of proper element calculations) the conversion from osculating to proper elements depends on $\omega$, $\Omega$, and the orbital properties of the rest of the planetary system. In these collisions, all of these quantities are nearly identical for all the simulated family members, suggesting that the shape and properties of the $a$-$e$-$i$ distribution is well preserved when converting from osculating to proper elements. To confirm this theoretical expectation, several 50 Myr integrations of synthetic families were completed in REBOUND, an accurate and efficient N-body integration package \citep{rein2012rebound}. From these integrations, proper elements were calculated and compared to the original distribution of osculating elements. In each case, all family members on non-resonant, stable orbits maintained their overall $a$-$e$-$i$ distribution shape. However, the 'center' of the proper element distribution was shifted from the center of the osculating element distribution, as would be expected. Since our model includes the center of the family as a free parameter and uses the shape of the $a$-$e$-$i$ distribution to constrain the models, the comparison of osculating elements to proper elements is not a major source of systematic uncertainty in our interpretation and results. 

We mention here that models which propose a dominant direction in the $\boldsymbol{\Delta v}$ distribution are essentially equivalent to shifting the orbital properties of the collision center. For example, \citet{2012MNRAS.419.2315O} propose an ejection field with $\Delta v$ of 400 m s$^{-1}$ in a particular direction with a dispersion around that position of 150 m s$^{-1}$. Such a model is effectively the same as an isotropic distribution of 150 m s$^{-1}$ around an apparent collision "center" that is 400 m s$^{-1}$ offset from the actual collision center. In combination with the issue of osculating vs. proper element comparison, this means that the collision "center" should be interpreted carefully. 

\subsubsection{Planar-Isotropic Model}

Simulations of \citet{2010ApJ...714.1789L} (see Figure \ref{planar_eje}) give a strong prediction that Haumea family members were ejected in a narrow plane. This is consistent with the physical ansatz that the family was created in order to shed angular momentum, leaving Haumea spinning near break-up. Indeed, any model that connects Haumea's spin to the formation of the family is likely to eject objects in a preferred plane, perpendicular to the extreme spin angular momentum. Other hypotheses that involve catastrophic collisions produce isotropic distributions. 

Preferring to have a model that can smoothly choose between being perfectly planar and completely isotropic, we develop a planar-isotropic (PI) model may yield different levels of "planarity", including an isotropic distribution. In order to capture a wide range of possibilities, we use a von Mises-Fisher distribution on a sphere (inspired by \citealt{2009ApJ...696.1230F}). A von Mises-Fisher distribution has both a concentration parameter, $\kappa$, and a mean direction parameter, $\boldsymbol{\mu}$, which we characterize by azimuth and altitude angles $\theta$ and $\phi$, respectively\footnote{The expression for the von Mises-Fisher distribution is: $f(\boldsymbol{x},\kappa,\boldsymbol{\mu}) = C(\kappa)\exp{(\kappa \boldsymbol{\mu}^{T} \boldsymbol{x})}$, where $C(\kappa)=\frac{\kappa}{4 \pi \sinh{\kappa}}$. This is, essentially, a normal distribution on the surface of a sphere.}. The greater the value of $\kappa$, the tighter the concentration of the distribution around $\boldsymbol{\mu}$. To create a planar distribution, random draws from the von Mises-Fisher distribution (for the proposed $\kappa$ value) were rotated a random amount in the plane perpendicular to $\boldsymbol{\mu}$. For high values of $\kappa$, the result is a planar ejection field, with larger values of $\kappa$ corresponding to more planar distributions as discussed in greater detail below. At $\kappa = 0$, there is no concentration around $\boldsymbol{\mu}$ and the von Mises-Fisher distribution becomes isotropic over the sphere (which is preserved by our random rotations). In the PI model, $\kappa$, $\theta$, and $\phi$ are all free parameters such that all distributions, from a completely isotropic distribution to a razor thin disk (with any orientation) are allowed.\footnote{There is technically a rotation degeneracy between $\Omega_c$ and $\theta$, but since we are not attempting to reconstruct the actual orientation, fixing $\Omega_c$ to 0$\degr$ does not affect fits to this model.}

\begin{figure*}
    \begin{center}
    \includegraphics[width=\textwidth,height=\textheight,keepaspectratio]{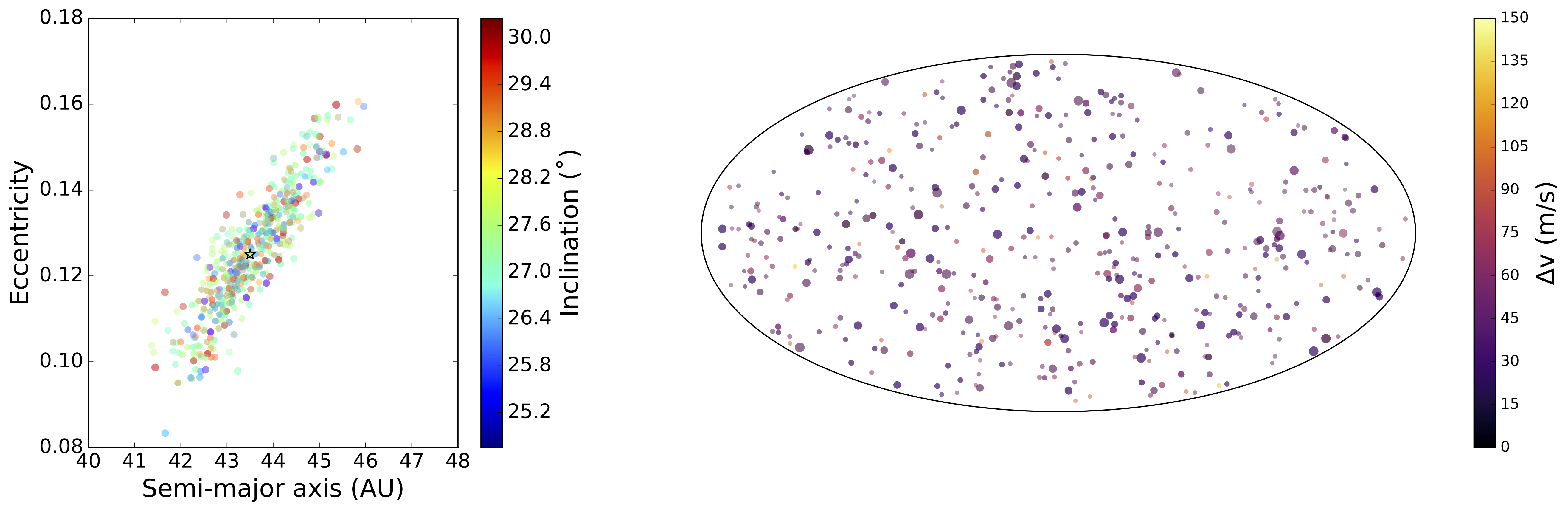}
    \caption{Ejection pattern of family members from an isotropic ejection in $a$-$e$-$i$-detectability-$\Delta v$ space (left panel) and the same family in a spherical equal area projection of the ejection vectors ($\Delta v_x$, $\Delta v_y$, $\Delta v_z$, right panel). A star shows the collision center; the collision center orbital elements are taken from \citet{2007AJ....134.2160R} and produce the distinct tilted elliptical shape in $a$-$e$ space, as is well known for collisional families. As described in the text, other parameters are used to specific the number-size-velocity distribution of simulated family members; here we have $\alpha = 0.2$, $S = 0.8$, $\beta = 0.1$, $k = 1.5$, and $\lambda = 1.7$. In the left panel the detectability of each particle is shown by transparency; high detectability as opaque and low detectability as transparent (e.g., at higher semi-major axis). The size of each point corresponds to the $\Delta v$ of that family member. In the right panel, a spherical equal area projection shows the direction and magnitude of each family member's ejection velocity. The directions are represented in the same way that a map of the Earth is projected onto a flat surface. The projection of the $\boldsymbol{\Delta v}$'s is done in such a way that any two regions of equal area on the flat projection cover the same area of the sphere. The size of each point corresponds to the radius of the simulated family members and the transparency to the detectability, as in the left panel. The right panel demonstrates the isotropic ejection pattern ($\kappa = 0.01$) as the $\boldsymbol{\Delta v}$'s are approximately equally distributed across the projected sphere, which is manifested in the orbital element space shown in the left panel as a high density region near the collision center, with a wide range of inclinations. Note that family members nearer the edges of the distribution tend to be less detectable, due to higher absolute magnitudes correlating with higher ejection velocities. }
    \label{isotropic_eje}
    \end{center}
\end{figure*}

\begin{figure*}
    \begin{center}
    \includegraphics[width=\textwidth,height=\textheight,keepaspectratio]{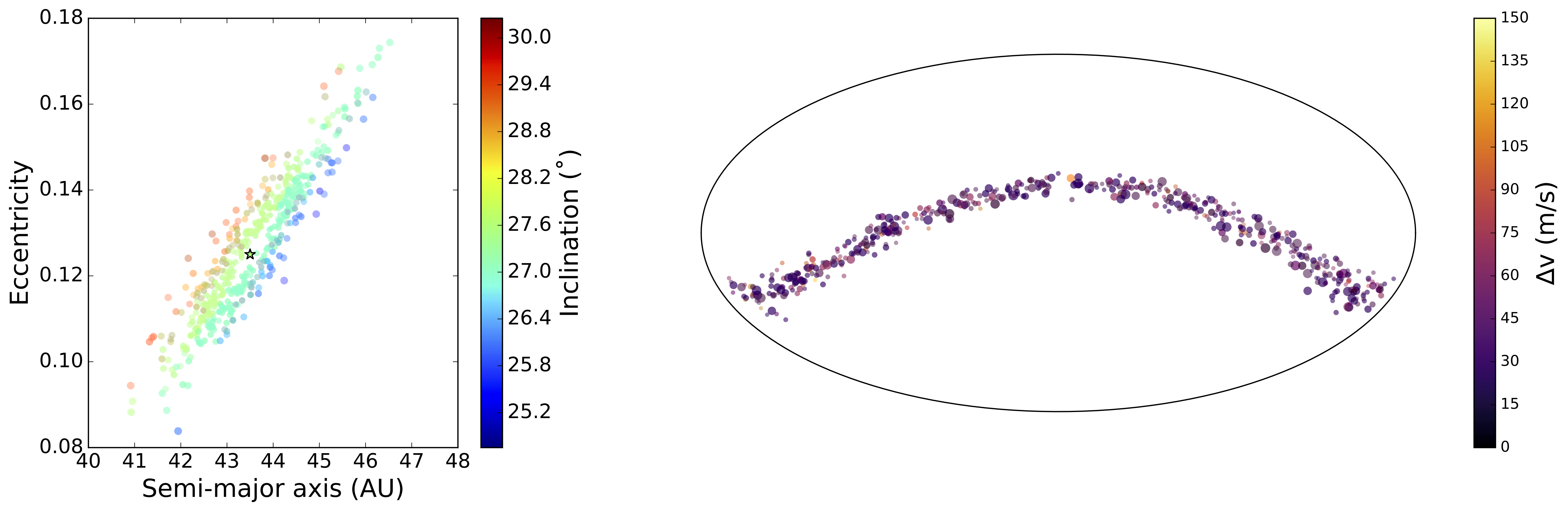}
    \caption{A graze-and-merge collision in the same style as Figure \ref{isotropic_eje} and with the same parameters except the family members are constrained to be ejected in a near planar distribution by setting $\kappa = 10^{3}$. As seen in the right panel, all family members are ejected in a plane. This ejection pattern results in a family where inclination is highly correlated with $a$ and $e$, producing the rainbow effect. The hole near the center of the distribution results from our $\boldsymbol{\Delta v}$ distribution having a minimum ejection. This correlation is to be expected; the ejection pattern is essentially two dimensional, embedding it into the three-dimensional $a$-$e$-$i$ distribution results in a correlation.}
    \label{planar_eje}
    \end{center}
\end{figure*}

This method gives a random distribution of distances and we set the magnitude of $\boldsymbol{\Delta v}$ equal to $\Delta v_{init}$ from Equation \ref{lykawka2}. As described above, we then calculate the resulting orbital elements. Figures \ref{isotropic_eje} and \ref{planar_eje} illustrate the $a$-$e$-$i$ distribution from this model in the isotropic ($\kappa \approx 0$) and planar ($\kappa = 10^3$) cases, respectively. This distribution can then be compared to the observations. With knowledge of the original distribution of $\boldsymbol{\Delta v}$, we can also show this distribution as illustrated by a spherical equal-area projection of the magnitude and direction of ejection velocities in these Figures.

The shape of the isotropic distribution is consistent with that found for asteroid families and \citet{2007AJ....134.2160R}. The planar ejection distribution shows a strong correlation between $a$, $e$, and $i$; this is expected because it is a fundamentally two-dimensional distribution embedded in three-dimensional orbital element phase space. As desired, the PI model prediction of a planar ejection distribution has strong predictions for the orbital distribution of Haumeans.  

\subsubsection{Delayed Ejection Planar-Isotropic Model}

Changes to the collision orbit $\omega_c$ and $M_c$ values are known to change the apparent $a$-$e$-$i$ distribution shape, e.g., it changes the orientation of the near-elliptical cloud of simulated family members. Previous models of asteroid families and the Haumea family have made a reasonable assumption that the timescale for ejecting family members is very short compared to Haumea's heliocentric orbital period ($\sim$300 years). However, a graze-and-merge collision leaves a significant fraction of material in an orbiting disk. Scattering interactions between objects in the disk could lead to ejections and these ejections may occur on timescales over which the effective mean anomaly $M$ of the collision orbit is actually changing. (The orbital precession timescale of millions of years indicates that keeping $\omega_c$ fixed is still reasonable.)

As a result, we propose a new variation on the PI model that we call "delayed ejection" planar-isotropic  (DEPI) model. This model uses the same algorithm to create the $\boldsymbol{\Delta v}$ distribution as before. However, to create a delayed ejection effect, rather than adding these velocities to Haumea's velocity at the time of impact, each family member was assigned a mean anomaly offset at which it was ejected. These offsets were chosen by drawing from an exponential distribution with scale parameter $\gamma$. In this case, $\boldsymbol{\Delta v}$ is added to Haumea's heliocentric velocity throughout its orbit, rather than at one point. 

\begin{figure*}
    \begin{center}
    \includegraphics[width=\textwidth,height=\textheight,keepaspectratio]{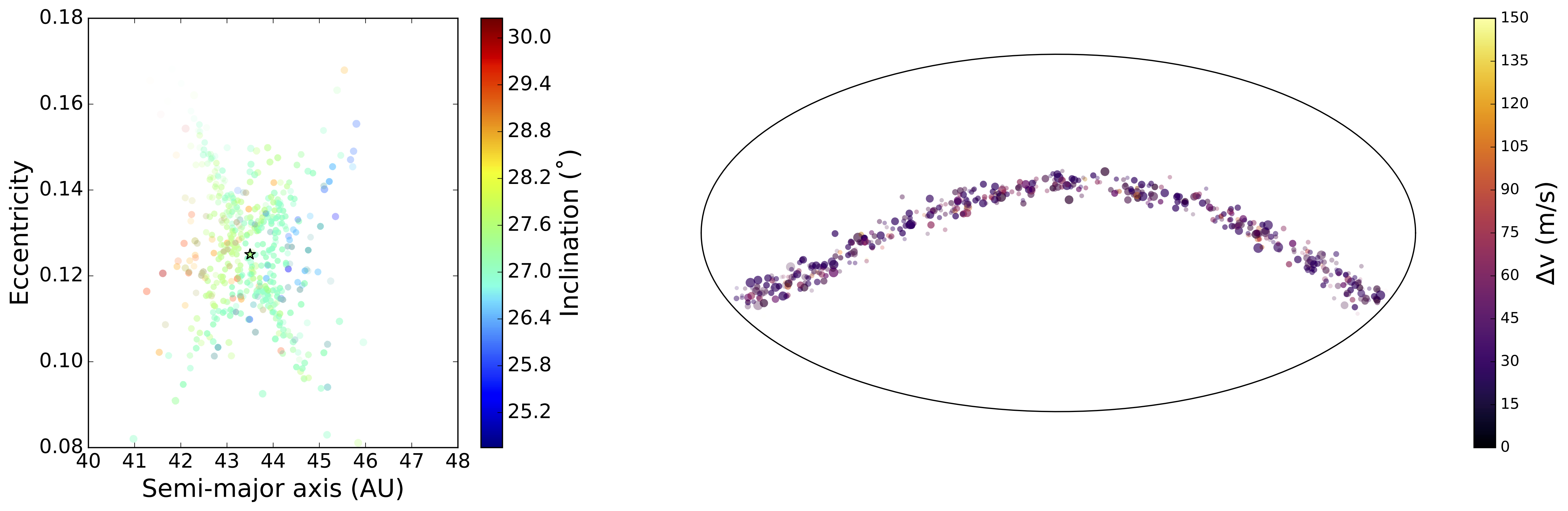}
    \caption{In the same style as Figure \ref{isotropic_eje}, a delayed ejection graze-and-merge collision is shown. In the right panel, the ejection directions are clearly identical to the ejection in a graze-and-merge collision (Figure \ref{planar_eje}). The first panel differs from that of the graze-and-merge ejection as the particles in a delayed ejection graze-and-merge collision are ejected over time, rather than all simultaneously, as in a graze-and-merge ejection. This emphasizes how the observed parameters (in the left panel) do not uniquely constrain the ejection directions without assuming (\emph{a priori} unknown) orbital angles of the collision, including the mean anomaly which is presumed to be changing in this delayed ejection model. This model changes the $a$-$e$ shape into an X-like shape, which results by averaging over the different orientations generated by the varying collision mean anomaly. Still, there are still strong correlations between $a$, $e$, and $i$ allowing delayed ejection graze-and-merge to be distinguished from any isotropic distribution.}
    \label{degam_eje}
    \end{center}
\end{figure*}

For $\gamma \sim 1$, most ejections happen within one orbital period and the resulting distribution is only slightly changed from the PI model. For $\gamma \gg 1$, ejections happen extremely quickly after the collision, resulting in a distribution nearly identical to that of the PI model. For $\gamma \ll 1$, ejections happen from random locations throughout the orbit. This is shown in Figure \ref{degam_eje}. As different values of $M_c$ result in different orientations of the ejection cloud in $a$-$e$-$i$ space, the resulting distribution is a superposition of these orientations, resulting in an X-like shape in $a$-$e$ space. 

\subsubsection{Satellite Collision Model}

In the satellite collision model, a satellite is destroyed catastrophically, resulting in an initial ejection field that is presumed isotropic. However, the heliocentric ejection velocities ($\boldsymbol{\Delta v}$) are affected in two ways: 1) the satellite's orbital velocity around Haumea gives all the ejected fragments a non-negligible velocity boost and 2) the gravitational influence of Haumea modifies their trajectories. 

In order to model these effects, we elect to continue using a very flexible model by including free parameters for the six satellite orbital elements ($a_{sat}$, $e_{sat}$, $i_{sat}$, $\omega_{sat}$, $\Omega_{sat}$, $M_{sat}$). This determines the position and velocity of the ur-satellite with respect to Haumea. $\Delta v_{init}$ from Equation \ref{lykawka2} is added to the satellite's Haumea-centric orbital velocity. The momentum of the impactor is neglected, but this only affects the results in the sense that the recovered satellite orbital elements could be incorrect. 

In this catastrophic collision, we assume isotropically distributed ejection directions. Using the initial ur-satellite position, we determine Haumea-centric orbital elements for each of the simulated family members. Objects that were bound to Haumea (or had a periapse distance within Haumea) were removed from the simulation. The remaining fragments, which were on hyperbolic orbits, had their escape direction and $v_{\infty}$ calculated. These velocities are then set to the heliocentric $\boldsymbol{\Delta v}$. 

\begin{figure*}
    \begin{center}
    \includegraphics[width=\textwidth,height=\textheight,keepaspectratio]{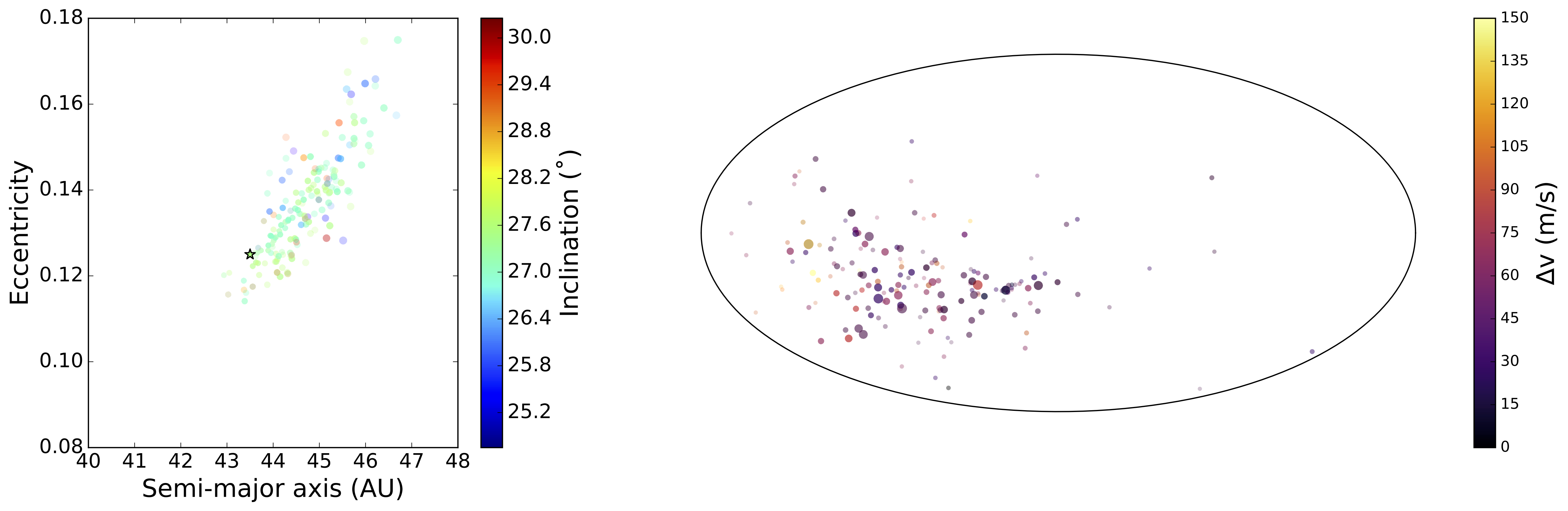}
    \caption{In the same style as Figure \ref{isotropic_eje}, a satellite collision ejection is shown. In the right panel, the ejection field clearly exhibits a preferred ejection direction, due to the initial velocity of the satellite; recentering on the middle of this distribution makes it appear much more isotropic. Additionally, outside of the main group of ejected family members, lie several objects that clearly deviate from the trend. This is due to Haumea's gravitational field bending the trajectories of those objects. In the left panel, there is a preferred ejection direction as well. With the collision center at $a = 43.5$ AU, $e = 0.125$, and $i = 27.50\degr$, the majority of the family is directed towards higher $a$ and higher $e$ regions. In practice, our model can recenter the family by shifting the heliocentric orbital elements of Haumea, allowing our model to focus on the re-centered shape of the ejection.}
    \label{satellite_eje}
    \end{center}
\end{figure*}

As mentioned above, the "boost" to the satellite's velocity is equivalent to shifting the velocity of the collision center's orbit. This can be seen in Figure \ref{satellite_eje} as an offset in the left panel and a concentrated distribution in the right panel. Considering the dispersion of the velocity distribution around the "mean" direction shows that the satellite collision appears much more isotropic. Depending on the properties of the satellite, there are some ejected family members that have their trajectories bent significantly causing a slight anisotropy in the ejection (see Figure \ref{satellite_eje}).

In our creation of satellite collision families, we approximated the escape trajectory of each family member by ignoring gravitational interactions with other escaping particles. This was done to remove the need for full n-body integrations, which are more accurate but not tractable. We tested this assumption by performing several n-body integrations, run in diverse regions of parameter space. These integrations were run in REBOUND, using the IAS15 integrator with an adjustable time step to resolve interactions between individual ejected particles \citep{rein2015ias15}. In the majority of cases, the n-body integrations showed little deviation from the 2-body approximation. In a few cases, several fragments had their trajectories significantly altered, creating a slightly different ejection pattern. However, our assessment is that the overall shape of the $a$-$e$-$i$-$H$ distribution would not be appreciably different from that of the 2-body assumption that we employ. 

\subsection{Detectability Cuts}

These models produce thousands of simulated family members, given specific values for all the input parameters. Before these can be fruitfully compared to observations, we need to remove objects that would not have ended up as detected objects in the Minor Planet Center. With the focus on comparing formation hypotheses, we performed two rough "detectability cuts": a cut based on dynamical stability and a cut based on observational detectability. Each cut gives a weight to each simulated object and these weights are multiplied together to provide a final "detectability" for each simulated family member that is used as a weight for that object in our later fits. 

The dynamical detectability estimate gave a weight of 0 to objects with perihelia <35 AU and a weight of 1 to objects with perihelia >38 AU. A simple linear slope was applied to particles with perihelia between 35 and 38 AU. This is meant to mimic long-term stability as calculated by \citet{lykawka2012dynamical} and \citet{2012Icar..221..106V}. 

Although observational surveys of the Kuiper belt are heterogeneous and far from complete, the fact that Haumea family members are, at present, randomly distributed in orbital angles while being in a concentrated region of $a$, $e$, and $i$ suggests that a detectability cut would affect most models in a similar way. Using each object's resulting orbital elements, an estimate of the orbit averaged absolute magnitude can be calculated. Based on the sensitivity of wide-field KBO surveys (e.g., \citealt{schwamb2010properties}, \citealt{2016arXiv160704895W}), we assign a observational detectability weight of 1 for objects brighter than 21$^{st}$ magnitude and a detectability weight of 0 for objects fainter than 23$^{rd}$ magnitude. A linear slope is again used between these limits. 

The results from the more accurate and precise detectability model of Pike et al. (2018, submitted) based on the Outer Solar System Origins Survey (OSSOS) are consistent with our results, suggesting that our crude observational completeness model is reasonable for our purpose of evaluating formation hypotheses. We also note here that Pike et al. (2018, submitted) did not find significant differences in the observational detectability of a planar model vs. an isotropic model. We feel confident that our detectability cuts are reasonable and do not bias us towards any of the tested models. 

\subsection{Observed Haumeans}
\label{sec:observed}
Simulated families can now be compared to the observed Haumeans, including several new detections since \citet{2007AJ....134.2160R}. In order to provide a clear comparison sample, we begin by selecting all known objects from the Minor Planet Center with osculating elements between $a$ = 40-47 AU, $e$ = 0.07-0.19, $i$ = 24$\degr$-33$\degr$. The proper elements of these 96 objects were calculated using very similar techniques to \citet{2007AJ....134.2160R}: determining the average orbital elements over a backwards integration of 50 MYr using the REBOUND integrator. Comparison of our results to the proper elements of AstDys\footnote{\texttt{\url{http://hamilton.dm.unipi.it/~astdys2/propsynth/tno.syn}}} showed excellent agreement. We used AstDys values for 2010 VR$_{11}$ and 2015 AJ$_{281}$ since these objects were not processed correctly by our pipeline. Our pipeline also focuses on producing accurate and precise uncertainties in the proper elements; we typically find very small variations (<1\%) compared to our expected systematic uncertainties and therefore these were ignored. 

We then limited our comparison to KBOs in a (proper) $a$-$e$-$i$-$H$ box with $a$ = 40-47 AU, $e$ = 0.08-0.17, $i$ = 26$\degr$-31$\degr$, and $H$ = 3.4-6.0 mags. These proper elements were chosen to encompass the range of orbital elements with $\Delta v \sim$ 150m s$^{-1}$, although there are regions of this space with $\Delta v >$ 150m s$^{-1}$. The H range was chosen to include the brightest known family member (2002 TX$_{300}$, $H$ = 3.4) at the lower end. The upper bound was chosen to limit background objects and was near the detectability cut limits. We also removed KBOs that showed evidence of resonance occupation, based on visual and statistical investigation of the orbital elements during the backwards integrations. 

The resulting list has 22 objects listed in Table \ref{tab:KBOs}. Family members from \citet{2007AJ....134.2160R} appear as do many new objects that are dynamically close to the family and can be considered good candidate Haumeans. Note that objects were neither included nor excluded based on color or spectral information. This ensures that our models are more robust to the "black sheep" hypothesis, which proposes that some dynamically nearby objects (like 2002 GH$_{32}$ and 2015 UQ$_{513}$) that do not share the same spectral/color features as Haumeans may still be collisional fragments \citep[e.g.,][]{carry2012characterisation}. 

\begin{figure*}
    \begin{center}
    \includegraphics[width=\textwidth,height=\textheight,keepaspectratio]{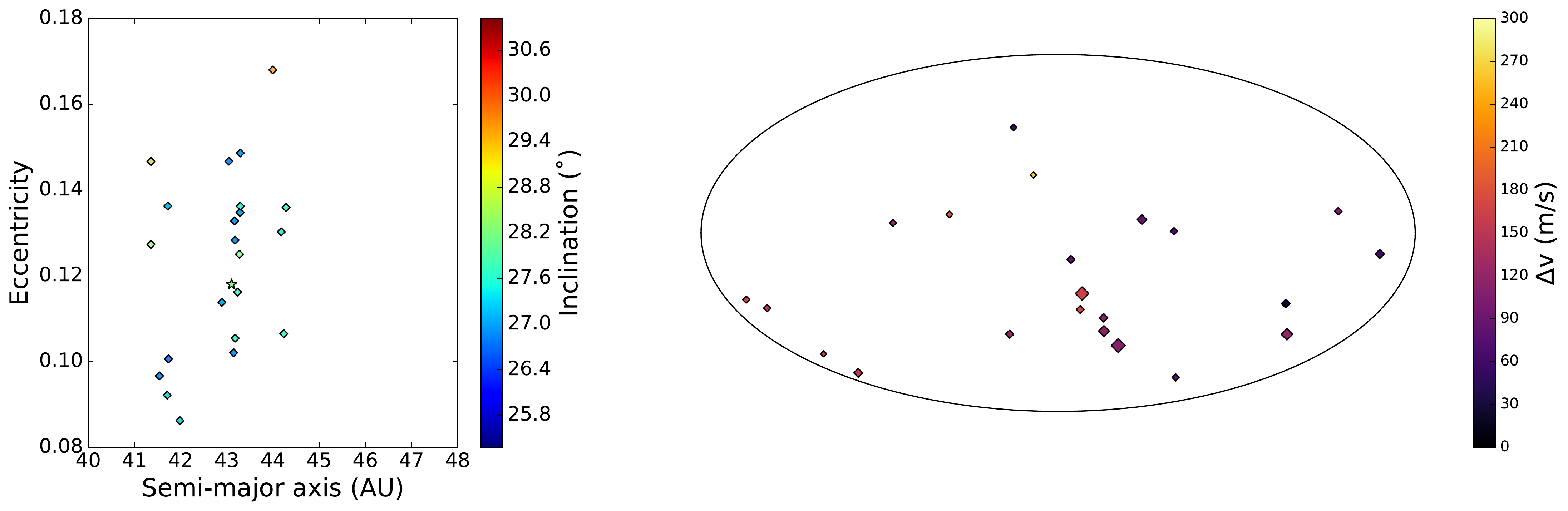}
    \caption{In the same style as Figures \ref{isotropic_eje}-\ref{satellite_eje}, the identified Haumeans visualized. In the 3-d spherical projection, the $\boldsymbol{\Delta v}$ is determined from the center of the family found in \citet{2007AJ....134.2160R}.}
    \label{fam_eje}
    \end{center}
\end{figure*}

\begin{table*}[t]
\centering
\caption{Data Set for Testing Family Models}
\label{tab:KBOs}
\begin{tabular}{cccccccc}
\hline
\hline
Number & Name & $a$ (AU) & $e$ & $i$ ($\degr$) & $H$ (mag) & $\Delta v$ (m s$^{-1}$) & Reference \\ \hline
55636 & 2002 TX$_{30}$ & 43.18 & 0.1283 & 26.97 & 3.4 & 108.3 & \cite{2007AJ....134.2160R}\\ 
202421 & 2005 UQ$_{513}$ & 43.29 & 0.1486 & 27.15 & 3.5 & 164.0 & \ldots\\ 
120178 & 2003 OP$_{32}$ & 43.14 & 0.1021 & 27.04 & 3.9 & 119.9 & \cite{2007AJ....134.2160R}\\ 
145453 & 2005 RR$_{43}$ & 43.16 & 0.1328 & 27.06 & 4.1 & 123.0 & \cite{2007AJ....134.2160R}\\ 
386723 & 2009 YE$_{7}$ & 44.28 & 0.1360 & 27.99 & 4.4 & 83.9 & \cite{trujillo2011photometric}\\ 
523645 & 2010 VK$_{201}$ & 43.18 & 0.1055 & 28.02 & 4.5 & 67.5 & \ldots\\ 
24835 & 1995 SM$_{55}$ & 41.73 & 0.1006 & 26.85 & 4.7 & 150.1 & \cite{2007AJ....134.2160R}\\ 
19308 & 1996 TO$_{66}$ & 43.23 & 0.1162 & 28.01 & 4.7 & 23.1 & \cite{2007AJ....134.2160R}\\ 
308193 & 2005 CB$_{79}$ & 43.28 & 0.1348 & 27.18 & 4.7 & 111.5 & \cite{schaller2008detection}\\ 
315530 & 2008 AP$_{129}$ & 41.72 & 0.1363 & 27.38 & 4.9 & 130.0 & \ldots\\ 
\ldots & 2014 HZ$_{199}$ & 43.04 & 0.1467 & 26.92 & 5.0 & 174.0 & \ldots\\ 
\ldots & 2015 AJ$_{281}$ & 43.35 & 0.1362 & 27.95 & 5.0 & 71.9 & \ldots\\ 
\ldots & 2014 QW$_{441}$ & 44.23 & 0.1065 & 27.98 & 5.2 & 106.9 & \ldots\\ 
416400 & 2003 UZ$_{117}$ & 44.18 & 0.1302 & 27.88 & 5.3 & 66.8 & \cite{schaller2008detection}\\ 
523627 & 2008 QB$_{43}$ & 41.71 & 0.0922 & 27.64 & 5.3 & 141.7 & \ldots\\ 
\ldots & 2014 LO$_{28}$ & 42.89 & 0.1138 & 27.26 & 5.3 & 74.0 & \ldots\\ 
\ldots & 2002 GH$_{32}$ & 41.98 & 0.0862 & 27.58 & 5.4 & 164.0 & \ldots\\ 
471318 & 2011 JF$_{31}$ & 41.35 & 0.1273 & 29.02 & 5.4 & 91.9 & \ldots\\ 
\ldots & 2010 VR$_{11}$ & 41.45 & 0.1467 & 29.36 & 5.6 & 181.0 & \ldots\\ 
471954 & 2013 RM$_{98}$ & 43.27 & 0.1250 & 28.78 & 5.7 & 14.9 & \ldots\\ 
\ldots & 2014 NZ$_{65}$ & 44.00 & 0.1680 & 29.80 & 5.8 & 263.0 & \ldots\\ 
523776 & 2014 YB$_{50}$ & 41.54 & 0.0967 & 27.01 & 5.9 & 133.6 & \ldots\\ \hline
\end{tabular}
\tablecomments{Starting with a large range of osculating elements, we calculated the proper elements of 96 KBOs using backwards integrations as in \citet{2007AJ....134.2160R}. All stable non-resonant objects with proper elements in the range 40 < $a$ < 47 AU, 0.08 < $e$ < 0.17, 26$\degr$ < $i$ < 31$\degr$ and 3.4 < H < 6.0 are listed here. These 22 KBOs include all the known non-resonance spectrally-confirmed family members, several new candidates, and likely several interlopers. References are given to justify the spectral confirmation of 8 KBOs. This list shows the observational constraints that our models are required to fit. The absolute magnitude $H$ is taken from the Minor Planet Center, which is known to have significant uncertainty. $\Delta v$ is calculated from the family center identified by \cite{2007AJ....134.2160R} and which is updated in Section \ref{sec:newhaumeans} below. }
\end{table*}

\subsection{Bayesian Parameter Inference}
\label{sec:bayesianfit}
Our goal now is to use infer the values of our multiple free parameters by comparing simulated families to the observational data. This is well suited to a Bayesian modeling approach, which we briefly describe here, referring the interested reader to the vast statistical literature on this subject

\subsubsection{Introduction to Bayesian Statistics}

Bayesian parameter inference estimates the joint probability distribution for a set of parameters ($\boldsymbol{x} \equiv a_c,\allowbreak e_c,\allowbreak i_c,\allowbreak \omega_c,\allowbreak M_c,\allowbreak \alpha,\allowbreak \beta,\allowbreak S,\allowbreak \lambda,\allowbreak k,\allowbreak \ldots$) conditioned on the observed data ($D$), written as $p(\boldsymbol{x}|D)$. Instead of a best fit and uncertainty, the "answer" is this probability distribution, called the \emph{posterior} since it is the result after fitting to the data. Looking at only one dimension at a time, the posterior probability distribution for each parameter can be summarized as a best fit with some uncertainty, but Bayesian parameter inference does not assume that parameters have Gaussian uncertainty and also fully characterizes correlations and covariances between different parameters. 
As is standard practice, we use Bayes Theorem to calculate the posterior distribution: 
\begin{equation}
    p(\boldsymbol{x}|D) = \frac{p(\boldsymbol{x}) p(D|\boldsymbol{x})}{p(D)}
\end{equation}
where $p(\boldsymbol{x})$ is the \emph{prior} probability distribution, $p(D|\boldsymbol{x})$ is the probability of getting the observed data assuming fixed parameters, and $p(D)$ is the unconditional likelihood of obtaining the data. The prior distribution represents one's belief about the parameters before using any data and therefore it makes sense to use very broad ranges that are "uninformative", i.e., they defer to the observational constraints in a way that does not introduce bias. The quantity $p(D|\boldsymbol{x})$ is called the likelihood, e.g., the likelihood of obtaining the observed data if the proposed model parameters were true. The denominator, $p(D)$ is called the "evidence" and requires finding the probability of finding each KBO in our data set over the entire range of model parameters (this integration is not shown explicitly in our equation for Bayes Theorm), which is computationally expensive. Since the evidence is not dependant on $\boldsymbol{x}$, we can ignore it as we only look at the relative posterior probabilities within each particular model.

Therefore, in our case, we are most interested in the prior distribution $p(\boldsymbol{x})$ and the likelihood $p(D|\boldsymbol{x})$. The prior distribution that we use for each parameter is given in Table \ref{tab:results} and the calculation of the likelihood is described below. Our parameters are generally well constrained by the data and we do not believe that our results are strongly influenced by our priors, as evidenced in tests of our methodology where the values for the priors did not change our results. We report our priors below so that future studies could redetermine the posterior distribution with different priors if desired. 

\subsubsection{Likelihood Calculation}

The central calculation of our model is a determination of the likelihood, $p(D|\boldsymbol{x})$. In our case, the data are $a$, $e$, $i$, and $H$ for each of the 22 KBOs. Each KBO is independent, so the total likelihood is the product of the individual likelihoods for each KBO (or, in practice, the sum of the logarithm of the individual likelihoods).

We use the above models with the proposed parameter values to generate $a$, $e$, $i$, and $H$ for $\sim$1000 simulated family members. To estimate the probability that a particular observed KBO would be produced by this simulated family, we approximate the $a$-$e$-$i$ distribution of the simulated family in 5 different $H$ bins by fitting the three-dimensional distribution of simulated orbital elements with a multivariate normal. Each $H$ bin has its own multivariate normal fit (characterized by a mean and a 3x3 covariance matrix) and these fits use the weight of each object from the detectability cut. The multivariate normal can then be used as a probability distribution, providing a likelihood for each observed object. 

Fitting a (weighted) multivariate normal is analytical and fast, but it is an approximation. However, inspection of the $a$-$e$-$i$ distribution in Figures \ref{isotropic_eje}, \ref{planar_eje}, and \ref{satellite_eje} shows that these can be reasonably approximated by Gaussian shapes, as the distribution of family members in each orbital element are approximately normally distributed.

In the case of the DEPI model, the synthetic families created displayed complex correlation in $a$-$e$-$i$ space, which may not be adequately described by a multivariate normal distribution. To create a likelihood function for this case, the number of synthetic family members near each tested KBO was counted and divided by the total number of synthetic particles similar to the method in \citet{brovz2018study}. For a sufficiently large number of simulated particles, this method reasonably estimates the likelihood of each KBO. For this calculation, we used $\sim$5000 simulated family members. As before, all KBO likelihoods were combined, giving a total likelihood for a given set of parameters. 

\subsubsection{Using a Mixture Model to Increase Robustness and to Model Interlopers}

With the use of multivariate normals, our likelihood calculations could be skewed by outliers. In addition, the majority of KBOs in our data bin are not confirmed family members and some of them are likely to be interlopers, e.g., objects randomly fell into that region of orbital element space and which are not related to the family (and therefore should not be used to determine the properties of the family). 

The effect of outliers and interlopers can be significantly mitigated by using a mixture model, where the observed distribution of KBOs is assumed to be a mixture of the simulated family and a background population model for the Kuiper Belt. We can include $f_{int}$, the fraction of interlopers in our population, as another free parameter and include it in the analysis. This implies that the likelihood of the mixture model is 
\begin{equation}
   \mathcal{L}_{mixture} = f_{int} \mathcal{L}_{int} + 
   \left( 1 - f_{int}\right) \mathcal{L}_{family} 
\end{equation}
where $\mathcal{L}_{family}$ is the likelihood that this object is a family member, calculated as described above and $\mathcal{L}_{int}$ is the likelihood that the object is an interloper, i.e., from the background population. 

We elected to use the parameterized empirical background population from the CFEPS-L7 model of the hot Kuiper belt \citep{petit2011canada}. Technically, CFEPS-L7 is describing osculating elements and not proper elements, but for the same reasons discussed above, this is not a serious issue. We elect to use the model as specified in \citet{petit2011canada} and to not introduce any additional free parameters in our model. Since this model has no free parameters, we can pre-compute $\mathcal{L}_{int}$ for all of our KBOs at their locations in $a$-$e$-$i$-$H$ space. Using the prescription for this model, we generate $10^7$ simulated background KBOs and apply our detectability cuts.\footnote{We note that the inclination distribution of "sin(i) times a Gaussian with width $\sigma$" popularized by \citet{brown2001inclination} is equivalent to a Rayleigh distribution with width $\sigma$ (when $\sigma$ is as small as observed for the hot KBO, as we confirmed explicitly). Rayleigh distributions are more well-defined and the natural expectation of inclination distributions of planetary populations, e.g., \citet{lissauer2011architecture}.} From this distribution, a method similar to the DEPI model likelihood function was followed: the background likelihood of each observed KBO was approximated as the sum of the weights of simulated objects in a bin around it (of size $\Delta a$ = 0.1 AU, $\Delta e$ = 0.01, $\Delta i$ = 0.5$\degr$, and $\Delta H$ = 1.0 mag) divided by the total of all the weights. 

The calculation of $\mathcal{L}_{int}$ focuses only on the $a$-$e$-$i$-$H$ distribution and not on the total number of objects. As the CFEPS-L7 model includes a normalization, we can estimate the total number of background objects that we expect among our KBOs. Evaluating the probabilities from \citet{petit2011canada} gave a total background population of $10_{-2}^{+2}$ objects. Using the average detectability of the background population, we estimate that there are $4_{-1}^{+1}$ background objects in our observations.

Our mixture model likelihood increases robustness to outliers of any kind because a KBO that doesn't fit the simulated family cannot severely penalize the overall likelihood. On the other hand, the mixture model assumes that all our KBOs have some probability of being background interlopers. However, at least 8 of these objects are spectrally confirmed family members (2002 TX$_{300}$, 2003 OP$_{32}$, 2005 RR$_{43}$, 2009 YE$_{7}$, 1995 SM$_{55}$, 1996 TO$_{66}$, 2005 CB$_{79}$ and 2003 UZ$_{117}$). To force the models to match these objects, the likelihoods for these objects are set equal to $\mathcal{L}_{family}$. 

We experimented with adding a free parameter for each individual KBO (that wasn't spectrally confirmed) that represented the probability that this individual object was a family member / interloper. While we had some success, this effectively doubled the parameters in our model, making it more difficult to find numerically and statistically stable solutions. The mixture model is effectively a marginalization of these individual parameters and we elected to use it. We discuss below how we use the values of $\mathcal{L}_{family}$ and $\mathcal{L}_{int}$ for each KBO to estimate the probability that it is a family member.

\subsubsection{Markov Chain Monte Carlo Sampling to Determine the Posterior Distribution}

A single likelihood evaluation (multiplied by the prior) only provides one piece of information on the posterior distribution. Obtaining the full posterior distribution requires calculating integrals over parameter space; in practice, these integrals are calculated numerically using Markov Chain Monte Carlo (MCMC) sampling. MCMC sampling can simultaneously explore parameter space, find the maximum likelihood (best fit), and determine the uncertainties in the parameters. The "burn in" is the exploration phase before MCMC finds and characterizes the best fit; after the burn in, MCMC samples are draws from the joint posterior distribution. For some angle parameters with symmetry, we found that choosing a smaller range to explore removed degeneracies; as a result, we do not expect our posterior distribution to be strongly multimodal with many statistically significant local maxima. 

With many correlated free parameters, our model required a non-trivial MCMC sampling algorithm (e.g., a method of choosing the next set of parameters given the properties of the current set of parameters). We used an Affine Invariant Ensemble Sampler (AIES) MCMC as implemented in \texttt{julia} with \texttt{AffineInvariantMCMC.jl}\footnote{\texttt{\url{https://github.com/madsjulia/AffineInvariantMCMC.jl/}}} which is based on the algorithm in \citet{goodman2010ensemble} and \citet{foreman2013emcee}. This MCMC sampler, unlike other samplers, evaluates probabilities at many locations at once, before taking its next step. These various locations are usually called "walkers." In each step, each walker uses the probability distribution found by the ensemble of walkers in the previous step to propose a new step. This allows the walkers to collectively explore parameter space more efficiently, when compared to standard MCMC algorithms. Additionally, since the sampler used is affine invariant, the sampler does well exploring parameter space with covariances among the parameters. 

For each model, 5000 starting positions for the walkers were selected at random from the prior distributions (where $\mathcal{L} \neq 0$). The samplers were then initialized and run. For each Markov Chain, the sampler ran for a 10$^5$ step burn-in. After the burn-in was discarded, the sampler ran for 10$^5$ additional steps.

Due to the nature of the AIES MCMC algorithm, when there are local maxima in the likelihood function, the algorithm's efficiency can be severely degraded. Testing showed decreased efficiency due to this problem, hence we implemented a clustering algorithm that can effectively identify and remove walkers which exhibit this issue, thereby increasing efficiency \citep{hou2012affine}.

\subsubsection{Model Selection}

Bayesian modeling includes powerful techniques for selecting between multiple models/hypotheses. However, these involve more complicated and larger calculations and may not be robust to our systematic errors. As a result, we use a simpler technique to give insight into how our models compared. 

In particular, we use the "Bayesian" Information Criterion (BIC). Even though the BIC is not actually Bayesian, this criterion is useful for comparing different models (it is similar in spirit to the reduced chi-square). Adding parameters always improves the likelihood, but this may not be statistically significant or meaningful. BIC takes this into account by penalizing models with more parameters: 

$$BIC = \ln{(n)}k - 2\ln{(\hat{\mathcal{L}})}$$

where $n$ is the number of observations or data points, $k$ is the number of parameters in the model, and $\hat{\mathcal{L}}$ is the best likelihood value given by the observations. A model with a larger BIC is disfavored. A more robust model comparison can be performed by including the calculation of the Bayesian evidence, but this is more complicated and not required to draw reasonable conclusions for our specific problem.

\subsection{Testing Our Fitting Routines}
\label{sec:testing}
To test the accuracy of each model and the fitting techniques used, we created artificial family member data to run tests on. This was done by outputting the $a$-$e$-$i$-$H$ data created by each model, from a specific set of model parameters. These families were specifically chosen as they were a archetypal example of each formation model. To keep the tests similar, many of the parameters (collision center orbital elements, $\Delta v$ distribution parameters, and $H$ distribution parameters) were held fixed. Along with data output by each model, several interlopers were also placed in the families to be tested, in an effort to validate our mixture model. Determining posterior distributions for these cases where we know "ground truth" allows us to assess the effectiveness of our techniques. 

Initially, artificial families were fit with the same model used to create the families. In these tests, the sampler correctly identified most important model parameters. 

For the PI model applied to a planar family, the sampler correctly found that the family exhibited a significant angle constraint with the correct orientation, correctly found the collision center, and identified that there were interlopers in the data set. This establishes that our technique can successfully identify families created by via a graze-and-merge collision. 

In addition, the PI model was also tested on several isotropic families. For these families, the sampler correctly found the collision center orbital elements. It did, however, find favorable solutions with $\kappa$ higher than the true value for these families ($\kappa = 5-10$ with $\kappa_{true} = 0.01$). The median value for the overall distribution was near the true value, but acceptable solutions with higher than expected $\kappa$ suggests that the PI model will overfit the data, finding a quasi-planar distribution even in an isotropic family. We suspect that this is a result of the small number of family members, the presence of interlopers, and the ability of the mixture model to "ignore" certain KBOs. We emphasize that this overfitting issue was restricted to the case where the true family was isotropic; planar families were accurately identified. 

For the DEPI model, the sampler correctly found the correct collision center orbital elements, the angle constraint with a reasonable orientation, and most importantly, found the correct $\gamma$, resulting in the posteriors having a significant delayed ejection effect. 

For the satellite collision model, the sampler correctly identified $a_{sat}$, $e_{sat}$, $i_{sat}$, $\Omega_{sat}$ and $M_{sat}$, although $a_{sat}$ and $e_{sat}$ had wide uncertainties. While it could reproduce the orbital elements of the collision center, these had uncertainties much wider than those of the other models. This can be explained by a degeneracy between the orbital elements of the satellite and collision center. These tests show that the sampling techniques used can also recover a satellite collision family. 

All of these tests showed several weaknesses in our models. Firstly, the tests showed that the samplers could not strongly constrain many of the number-size-velocity distribution parameters (e.g., $\alpha$, $S$, $\beta$, $\lambda$, and $k$). While the correct values of each parameter were usually within the 1$\sigma$ range of the posteriors, the posteriors were frequently broad, giving a wide range of acceptable values. While this is obviously a limitation of our model, the main objective of model selection does not require tightly constrained posteriors on each parameter. In fact, the large range of possible values allows us to be more confident that the performance of a model is not systematically biased by an assumed number-size-velocity distribution. 

When families with large interloper fractions ($f_{int} \approx 0.4$) were tested, the sampler produced poor results with the same computation time as the other tests. With extended computation time, it was possible to recover some model parameters. 

After the artificial families were tested with their own model samplers, we "cross-checked" the models. In these, each type of family was tested with each model and the results were compared across models using the $\Delta$(BIC).

For the planar family, the PI model had a 150 $\Delta$BIC advantage over the satellite collision model. This is a very strong result, indicating that the PI model sampler can produce far better fits than the satellite collision model on a planar family, as expected. 

For the isotropic family, each model sampler provided a reasonably good fit to the data, with the PI model having a modest 5 $\Delta$BIC advantage over the satellite model. While the PI model was used to create the isotropic families tested, the satellite model also has the capability of effectively producing isotropic families. It can do this by increasing $a_{sat}$ to the point where a collision can take place when the satellite is far from Haumea, reducing its Haumea-centric velocity and minimizing the effects of Haumea's gravity on each particle. 

For the satellite collision family, the satellite model had a 6 $\Delta$BIC advantage over the PI model. While this is not a particularly strong rejection, the fit did clearly indicate a preference towards the satellite model, even after the penalties for a greater number of parameters.

For the DEPI model family, the DEPI model exhibited a vastly better fit to the data than any other model. This resulted in a $\Delta$BIC advantage of 86 over the PI model and 90 over the satellite model. This constitutes a strong rejection of the PI and satellite models, showing that the DEPI model is best at identifying DEPI families.

\begin{figure}
    \centering
    \includegraphics[width=\textwidth,height=\textheight,keepaspectratio]{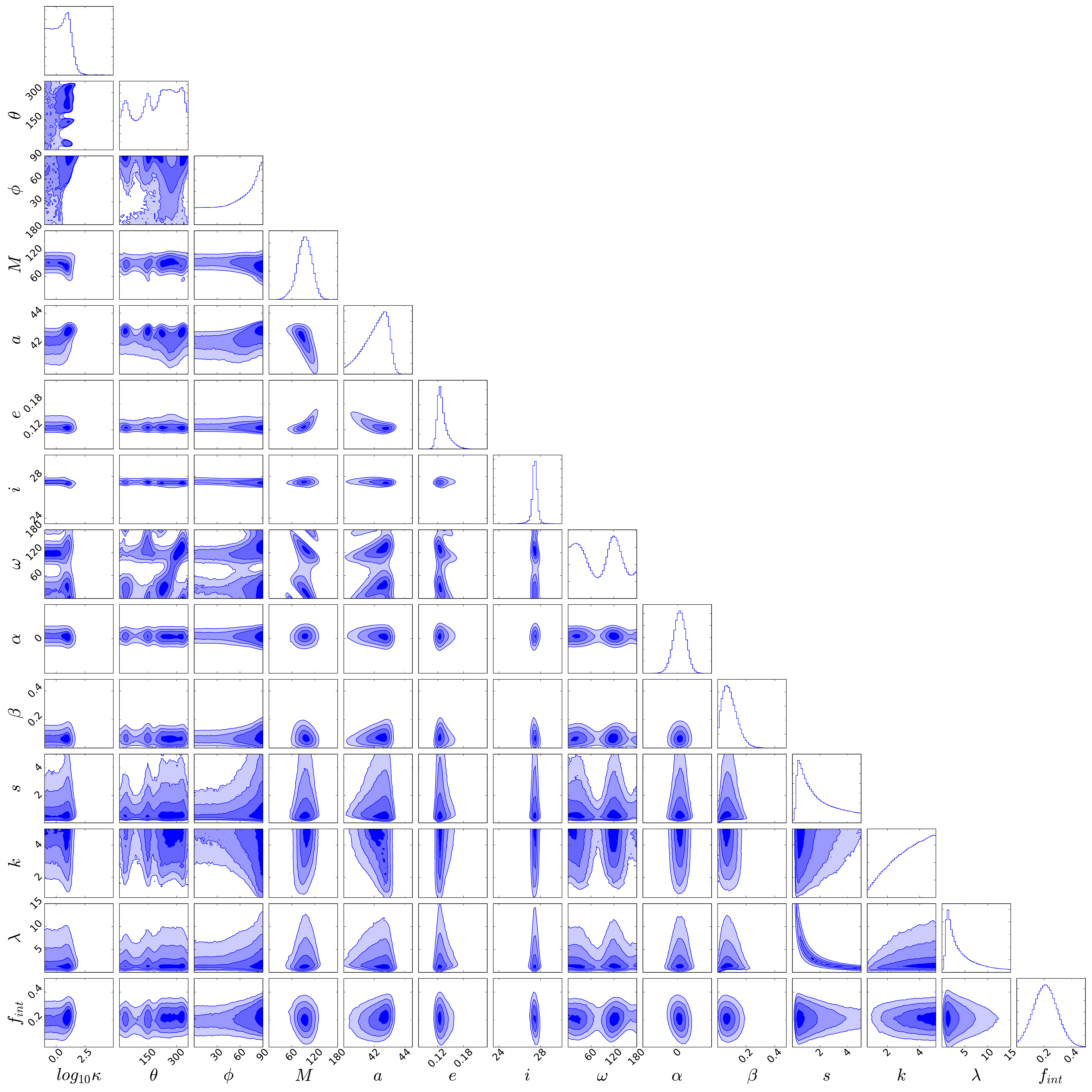}
    \caption{The "corner" plot for the PI model showing the posterior probability distribution for all our parameters (histograms) as well as joint probability distributions between every pair of parameters (contour plots). Results for the posterior distributions are also given in Table \ref{tab:results}. Of particular interest is the lack of support for large $\log(\kappa)$ values shown in the first column that would correspond to a planar ejection field; the graze-and-merge models of \citet{2010ApJ...714.1789L} would predict $\log \kappa \approx 2.5$. An quasi-planar ($\log \kappa \approx 1.2$ implying an inclination width of $\sim$60$\degr$) is slightly favored, but an isotropic model cannot be ruled out. The collision center orbital elements are similar to those seen in \citet{2007AJ....134.2160R}. The size distribution favors a shallow slope (as in Pike et al. 2018, submitted) with wide ranges of the other parameters. This mixture model coefficient $f_{int}$ indicates that roughly 0.2$\pm$0.1 of the KBOs in Table \ref{tab:KBOs} are interlopers.}
    \label{fig:pim_corner}
\end{figure}

\begin{figure}
    \centering
    \includegraphics[width=\textwidth,height=\textheight,keepaspectratio]{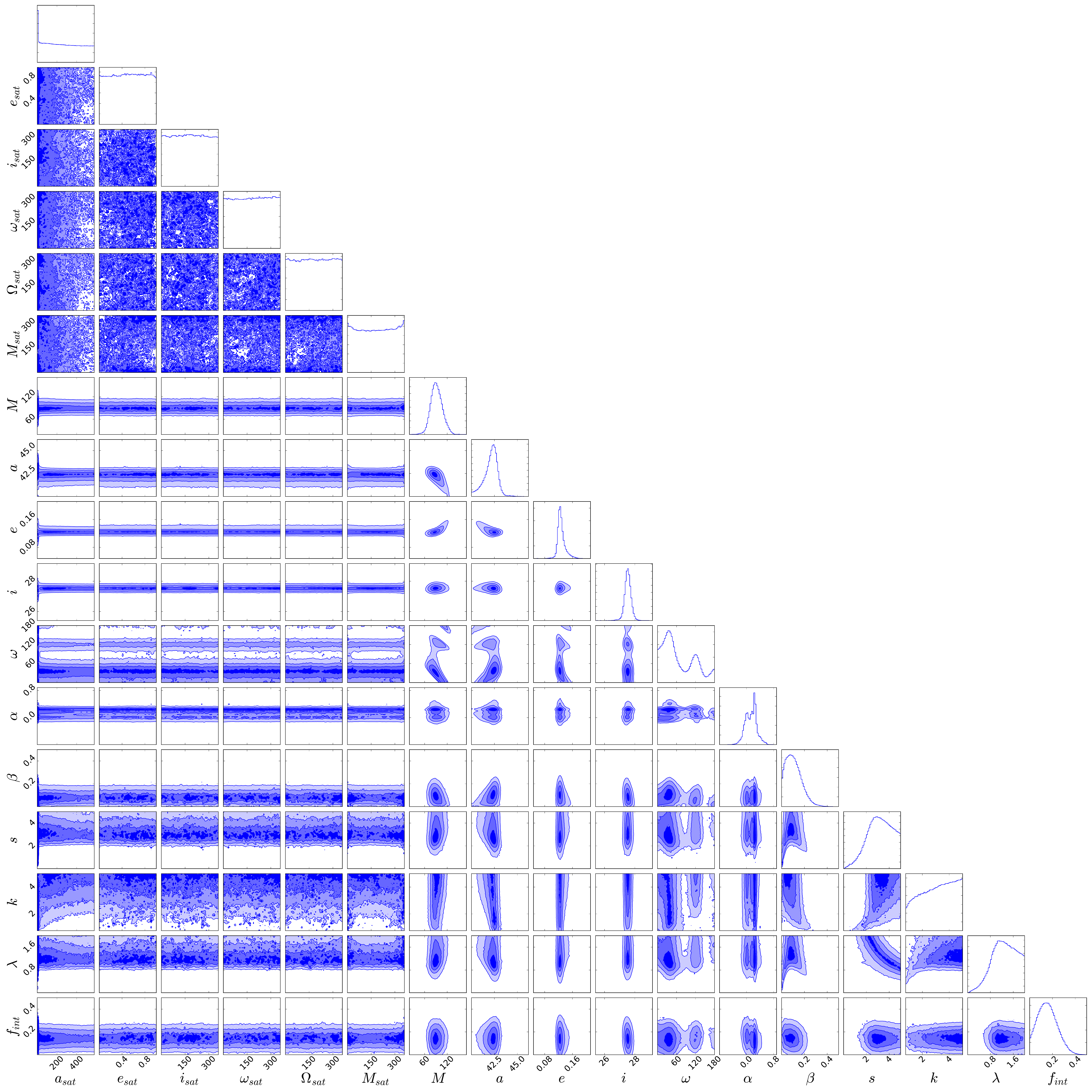}
    \caption{Similar to Figure \ref{fig:pim_corner} but for the parameters of the satellite collision model. A wide range of satellite orbital parameters are considered, but none are particularly favorable. Recasting these as the satellites Haumea-centric orbital velocity shows a strong preference for small velocities where this model becomes degenerate with the isotropic result from PI model. This model also favors a smaller $\lambda$ which implies a smaller and tighter $\Delta v$ distribution.}
    \label{fig:sat_corner}
\end{figure}
 
\section{Results}
\label{sec:results}
Including testing and many intermediate analyses, well over $10^9$ different model evaluations were performed.\footnote{The authors are willing to share intermediate results and analysis software upon request.} The results that we present here are from "final" runs with 5000 walkers and 10000 links. Convergence was confirmed by visual inspection of plots of parameter values vs. generation, parameter values vs. likelihood, and by the smoothness of the individual (marginal and joint) posterior distributions. The results of these runs are summarized by 1$\sigma$ confidence intervals in Table \ref{tab:results} and by Figures \ref{fig:pim_corner}-\ref{fig:sat_corner} showing a "corner plot" \citep{cornerplots} which shows the posterior distribution for each individual parameter and the joint distribution as a contour plot. 

\begin{table*}[t]
    \centering
    \caption{Results of the MCMC Testing}
    \begin{tabular}{ccccc}
\hline
\hline
Model Parameter & Prior Distribution & PI Model & DEPI Model & Satellite Collision Model \\ \hline
Fisher Distribution Concetration ($\kappa$) & $U(10^{-1},10^{5})$ & $2.44^{+12.08}_{-2.16}$ & $0.74^{+1.88}_{-0.53}$ & \ldots \\ 
Plane Azimuth ($\theta$) ($\degr$) & $U(0,360)$ & $211^{+98}_{-143}$ & $195^{+99}_{-115}$ & \ldots \\ 
Plane Altitude ($\phi$) ($\degr$) & $U(0,90)$ & $61^{+22}_{-40}$ & $41^{+29}_{-26}$ & \ldots \\ 
Exponential Distribution Scale ($\gamma$) & $U(0,100)$ & \ldots & $15.6^{+16.3}_{-10.0}$ & \ldots \\ 
Satellite Semi-major axis ($a_{sat}$) (10$^6$ km) & $U(10^{-3}, 10^{4})$ & \ldots & \ldots & $2.63^{+2.12}_{-1.96}$ \\ 
Satellite Eccentricity ($e_{sat}$) & $U(0,1.0)$ & \ldots & \ldots & $0.5043^{+0.3357}_{-0.3410}$ \\ 
Satellite Inclination ($i_{sat}$) ($\degr$) & $U(0,360)$ & \ldots & \ldots & $179^{+122}_{-121}$ \\ 
Satellite Argument of Periapse ($\omega_{sat}$) ($\degr$) & $U(0,360)$ & \ldots & \ldots & $182^{+121}_{-124}$ \\ 
Satellite Longitude of Ascending Node ($\Omega_{sat}$) ($\degr$) & $U(0,360)$ & \ldots & \ldots & $181^{+122}_{-123}$ \\ 
Satellite Mean Anomaly ($M_{sat}$) ($\degr$) & $U(0,90)$ & \ldots & \ldots & $182^{+124}_{-127}$ \\ 
Collision Semi-major Axis ($a_c$) (AU) & $U(40, 50)$ & $42.21^{+0.64}_{-0.99}$ & $42.07^{+0.44}_{-0.46}$ & $42.22^{+0.54}_{-0.82}$ \\ 
Collision Eccentricity ($e_c$) & $U(0,0.3)$ & $0.1282^{+0.0149}_{-0.0078}$ & $0.1254^{+0.0054}_{-0.0045}$ & $0.1258^{+0.0115}_{-0.0067}$ \\ 
Collision Inclination ($i_c$) ($\degr$) & $U(0,45)$ & $27.44^{+0.22}_{-0.21}$ & $27.64^{+0.17}_{-0.17}$ & $27.54^{+0.20}_{-0.19}$ \\ 
Collision Argument of Periapse ($\omega_c$) ($\degr$) & $U(0,90)$ & $91^{+48}_{-70}$ & $84^{+44}_{-54}$ & $50^{+74}_{-31}$ \\ 
Collision Mean Anomaly ($M_c$) ($\degr$) & $U(0,90)$ & $94^{+18}_{-19}$ & $83^{+12}_{-14}$ & $91^{+16}_{-14}$ \\ 
Absolute Magnitude Distribution Slope ($\alpha$) & $U(-1,1)$ & $0.05^{+0.15}_{-0.15}$ & $0.29^{+0.17}_{-0.17}$ & $0.15^{+0.13}_{-0.18}$ \\ 
Size-$\Delta v$ Correlation Constant ($\beta$) & $U(10^{-4},0.5)$ & $0.08^{+0.06}_{-0.04}$ & $0.12^{+0.08}_{-0.06}$ & $0.10^{+0.08}_{-0.06}$ \\ 
Size-$\Delta v$ Constant ($S$) & $U(10^{-4},5)$ & $1.45^{+1.95}_{-0.90}$ & $1.74^{+1.51}_{-0.90}$ & $3.20^{+1.14}_{-1.10}$ \\ 
Weibull Distribution Shape Parameter ($k$) & $U(0.75,5.0)$ & $3.48^{+1.07}_{-1.39}$ & $4.11^{+0.59}_{-0.82}$ & $3.18^{+1.26}_{-1.51}$ \\ 
Weibull Distribution Scale Parameter ($\lambda$) & $U(10^{-4},15.0)$ & $2.99^{+4.47}_{-1.72}$ & $2.17^{+2.12}_{-1.06}$ & $1.29^{+0.46}_{-0.43}$ \\ 
Interloper Probability ($f_{int}$) & $U(0,0.5)$ & $0.21^{+0.09}_{-0.09}$ & $0.09^{+0.06}_{-0.05}$ & $0.15^{+0.09}_{-0.08}$ \\ \hline

\end{tabular}

    \tablecomments{Shown here are the prior distributions (from which walkers are initialized) along with the posterior distributions of the parameters for each model. As described in Section \ref{sec:methods}, some parameters are shared among the models while others are unique. The uncertainties shown correspond to the 16$^{\textrm th}$ and 84$^{\textrm th}$ percentiles. The shape of the posterior distributions and correlations between them are shown in Figure \ref{fig:pim_corner} for the PI model and \ref{fig:sat_corner} for the satellite collision model. The full posterior distribution can be made available upon request. Note that the collision center orbital elements are compared to proper elements and are likely shifted from the original osculating elements. $U(a,b)$ refers to a uniform distribution from $a$ to $b$.}
    \label{tab:results}
\end{table*}

\subsection{PI model Results}

The PI model posteriors show that the Haumea family does not show significant ejection angle restriction, as would be expected from a graze-and-merge type collision. For relatively concentrated ejection fields ($\kappa \gg 1$), the von Mises-Fisher distribution is well approximated by a Rayleigh distribution with a width of $\sigma = \kappa^{-1/2}$. Therefore, we can convert $\kappa$ to a typical inclination ($\sigma$) under the caveat that this is only meaningful for nearly planar distributions. 

A graze-and-merge family is expected to have an ejection angle distribution of $\sigma_{median}\approx 5\degr$ \citep{2010ApJ...714.1789L} corresponding to $\kappa \approx 250$. This is ruled out by the posterior which gives $\sigma = 42.3_{-25.5}^{+84.7}$, with uncertainties indicating the 16$^{\textrm th}$ and 84$^{\textrm th}$ percentiles and where the large values imply isotropic distributions. 

\begin{figure}
    \centering
    \includegraphics[width=\textwidth,height=\textheight,keepaspectratio]{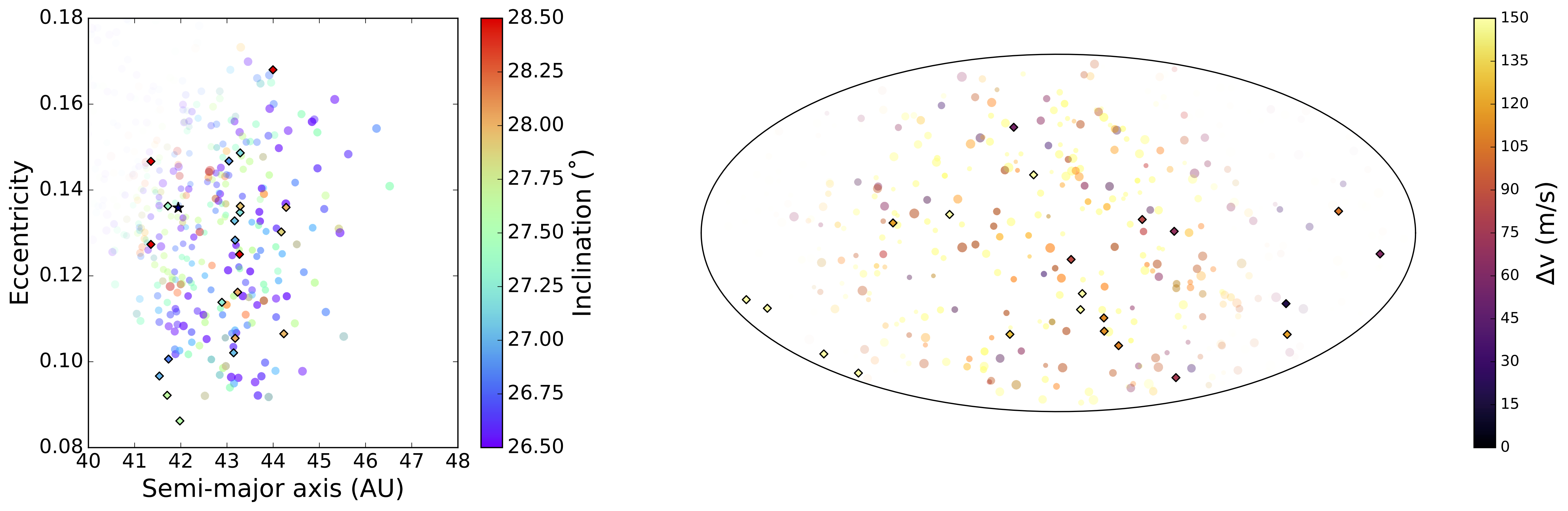}
    \caption{Shown, in the style of Figure \ref{isotropic_eje}, is an example of a synthetic family drawn from the posterior distribution of the PI model. The left panel demonstrates how the $a-e-i$ distribution is an adequate fit, while the right panel shows how this model is essentially isotropic. A planar model is clearly disfavored. Note that the left and right edges of the right panel show less simulated family members due to a lower detectability rather than a directional ejection. Family members in this region were ejected into unstable orbits, all of which have low detectability.}
    \label{fig:pim_bestfit}
\end{figure}

This angle constraint shows possible anisotropy in the distribution of KBOs near Haumea (see Figure \ref{fig:pim_bestfit}), which we will refer to as "quasi-planar". However, it is possible that this anisotropy is due to overfitting and not to the actual distribution of Haumeans. When this model was tested on families that were known to be truly isotropic (see Section \ref{sec:testing}), this quasi-planar result was favorable, although it was not the best fit. We suggest that the anisotropy suggested by our best fit is at least partially due to overfitting. Furthermore, our posterior distributions show that essentially isotropic ejection fields are adequate fits to the observed distribution of Haumeans. 

As a result, the distribution of $\boldsymbol{\Delta v}$ is not consistent with the graze-and-merge formation hypothesis of \citet{2010ApJ...714.1789L}. Indeed, any model that produces a planar ejection field is not consistent with these models. This truly provides a puzzle since it is difficult to imagine a single physical process that can simultaneously produce a near-breakup angular momentum for Haumea and a near-isotropic distribution of family members. 

\subsection{DEPI model Results}

The DEPI model was an attempt to match the near-isotropic distribution of Haumeans with a graze-and-merge formation hypothesis. Our detailed fits gave similar results to the PI model because the new key parameter $\gamma$ that describes exponential decay timescale of ejections was consistent with an ejection timeframe much smaller than Haumea's orbital period. In this regime, the two models are degenerate. At small timescales, like the posteriors require, the family is formed with the majority of particles being ejected quickly, creating a family shape similar to a graze-and-merge family, with a few particles being ejected at much later times. This can create a situation in which most of the initially ejected particles can fit the majority of KBOs in the data set, and particles ejected at later times matching only a few data points not covered by the initial family. Consequently, this can create a family which adequately covers all of the KBOs in the data set, reducing the need for objects to be ruled interlopers. This is indicated by the slightly lower $f_{int}$ for this model compared to the planar model. 

Testing and exploration indicated that ejection delayed over longer timescales were not supported by the data. Indeed, the characteristic X-shape and $a-e-i$ correlations for this model shown in Figure \ref{degam_eje} do not appear to describe the observations. 

\subsection{Satellite Collision Model}

The posterior distributions of the satellite orbital elements, as seen in the satellite model, show near uniform sampling from the prior distributions. This indicates that there were very few, if any, solutions found which closely matched the family distribution (see Table \ref{tab:results}).

\begin{figure}
    \centering
    \includegraphics[width=\textwidth,height=\textheight,keepaspectratio]{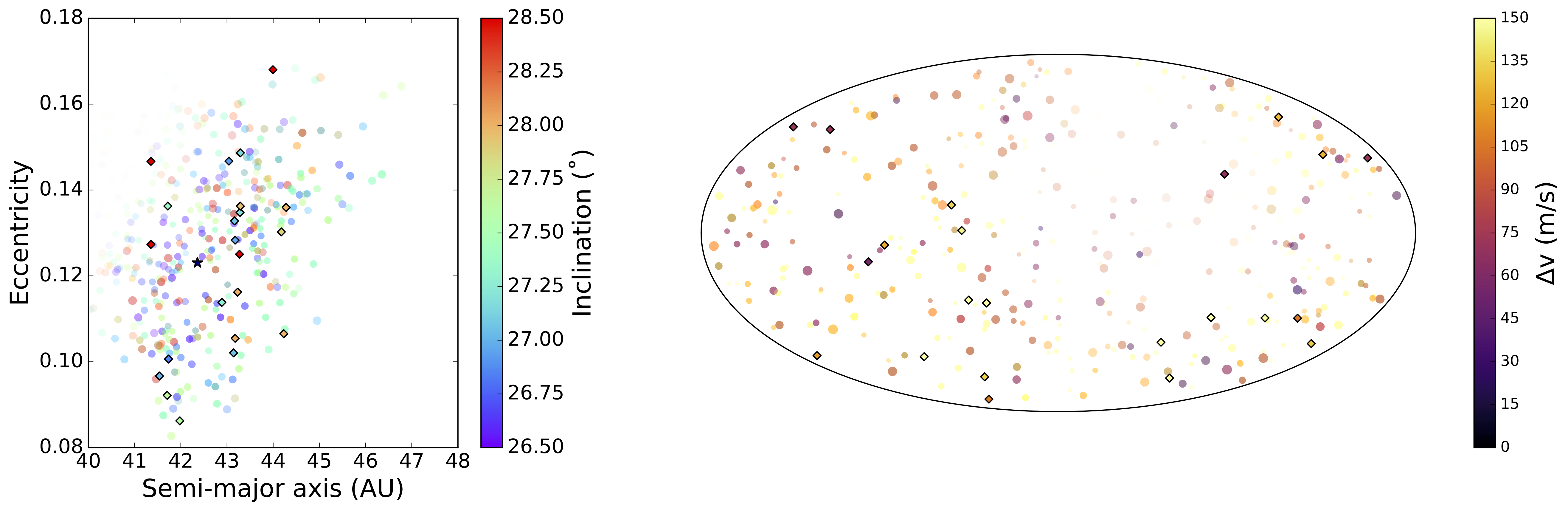}
    \caption{Shown is a synthetic family created from a draw from the satellite collision model posterior distribution, in the style of Figure \ref{isotropic_eje}. This synthetic family is clearly very similar to the synthetic family created in Figure \ref{fig:pim_bestfit}. This indicates ejections with similar model parameters and a satellite chosen to have a minimal boost and interaction with Haumea, unlike the proposed ur-satellite. Model comparison using $\Delta BIC$ clearly prefers the PI model to the satellite collision model.}
    \label{fig:sc_bestfit}
\end{figure}

As discussed above, the two primary deviations of the satellite model from an isotropic distribution are the velocity boost from the satellite's orbital velocity and the removal of objects that are bound (or impact) Haumea, which provides some anisotropy. Using the posterior distribution of the satellite orbital elements to infer the satellite's Haumea-centric velocity and position show that the fit clearly preferred very small velocities ($\sim 10$ m s$^{-1}$) and distant satellites. In this case, the differences between the satellite model and PI model are minimal and this model recovers the same solution as the PI model. This is clear in Figure \ref{fig:sc_bestfit}. Clearly, the ejection field is nearly isotropic. The posterior distribution is inconsistent with the expected velocities and distances of the satellite model (\citealt{2009ApJ...700.1242S}, \citealt{cuk2013dynamics}). 

The satellite model has a large parameter space with potentially narrow global minima. It is impossible to assert that we completely explored parameter space, but we feel that the inability of the satellite model to match the data is an indicator that it is not a viable explanation for the Haumea family. For example, our earlier tests showed that the shape of a collision formed by a satellite could be recovered. Furthermore, the significant velocity boost that would be expected from the putative ur-satellite could displace the center of the family enough that it would be difficult to explain how Haumea ended up in the 12:7 resonance.  

\subsection{Model Comparisons}

When each fit had its BIC calculated, the DEPI model had the highest BIC with a $\Delta$BIC of 70 over the PI model, and a $\Delta$BIC of 90 over the satellite model. While this indicates that the DEPI model is the model which can best describe the KBO data set, the issues of overfitting and physical mechanisms need to be taken into account. Although BIC attempts to resolve overfitting by penalizing models with more parameters, DEPI model has only 1 additional parameter ($\tau_{eje}$) which is in the regime where the DEPI model is hardly distinguishable from the PI model. We attribute the large $\Delta BIC$ value (and the tighter posteriors seen in Table \ref{tab:results}) to the larger number of simulated family members generated for these models and/or to the use of "voxels" for calculating likelihoods which are more accurate than the multi-variate normal. 

Furthermore, since the ejection distribution was quasi-planar/isotropic, the best-fit DEPI model is not consistent with the graze-and-merge formation hypothesis. This best-fit model is probably preferred because of overfitting or some process that slightly modifies the shape of the actual Haumeans that we are not including (like weak resonances). Or perhaps the true formation process did take several years to eject objects (in isotropic directions) as suggested by this model. 

The satellite model is disfavored by $\Delta BIC \simeq$ 20 compared to the PI model (which uses the same number of simulated family members and the multivariate normal likelihood estimation). Such a large $\Delta BIC$ strongly favors PI (somewhat similar to a 20-$\sigma$ result) and is another indictment against the satellite collision hypothesis. 

With these caveats in mind, we elect to use the PI model as our preferred final model; the resulting quasi-planar/isotropic posterior distribution from the PI model is taken as an empirical distribution of Haumeans for additional analyses below. 

\subsection{Family Member Probabilities and $\Delta v$'s}

\label{sec:newhaumeans}

Our model flexibly includes the possibility of a background interloper population, but does not explicitly identify which objects could be interlopers. After adopting a model, we can use the final posterior distribution to evaluate for each individual KBO the likelihood of belonging to the family or the background. This allows us to identify which KBOs are most likely family members and provides a way to aid in future identification of family members. 

To calculate the probability of each KBO in the test range, 10$^5$ random draws from the joint posterior were taken (e.g., self-consistent parameter sets from 10$^5$ random steps were taken). For each of these draws, we can use an odds ratio to determine the probability that it belongs to the family, $p_{fam}$: 

\begin{equation}
    p_{fam} = \frac{((1 - f_{int})/f_{int})(\mathcal{L}_{family}/\mathcal{L}_{int})}{((1 - f_{int})/f_{int})(\mathcal{L}_{family}/\mathcal{L}_{int}) + 1}
\end{equation}

The distribution of $p_{fam}$ over the 10$^5$ draws gives the posterior probability distribution that this KBO is a family member. In addition to this, the $\Delta v$ of each object can be found for each posterior draw as well (noting that the parameters include the collision center), resulting in a posterior distribution of $\Delta v$ for each KBO. Both of these measurements are compiled in Table \ref{familyness}. We used simulated families and other checks to validate this methodology.

The result is we identify 15 family members with median family probabilities greater than 0.9, which we refer to as "dynamically-confirmed" Haumeans. This includes the eight spectrally-confirmed family members, but since these were not allowed to be interlopers in the actual fit, the family probabilities shown here are subject to circular reasoning; the high probability values for these cases show that the family fit did not deviate much from fitting the spectrally-confirmed Haumeans. The new dynamically confirmed family members are 2005 UQ$_{513}$, 2010 VK$_{201}$, 2015 AJ$_{281}$, 2008 AP$_{129}$, 2014 LO$_{28}$, 2014 HZ$_{199}$, and 2014 QW$_{441}$. 

KBO (202421) 2005 UQ$_{513}$ is an interesting case: its proper elements somewhat near the center of the family and low $H$ push the family probability to 0.996. However, it is known to have a red optical slope \citep{pinilla2008visible} and low albedo \citep{vilenius2018tnos} and is thus much larger than all other family members (radius = 498 km from \citealt{vilenius2018tnos}). This probability is over-represented due to our assumptions about albedo. Neither our family models, nor the background model allow for different albedos. Future work may be able to add this to a model in order to achieve better solutions. Makemake also has a low $\Delta v$, but is not considered to be part of the family due to its size. 2005 UQ$_{513}$ may also be included in this category. 

As shown in Table \ref{familyness}, objects with low absolute magnitudes tend to be favored over objects with higher absolute magnitudes. This is due to the shallower absolute magnitude distribution of the family compared to the background, as the family contains many bright objects. Additionally, the objects with higher $\Delta v$ tend to have lower family probabilities. Since the family is so compact, the family members outside of this densely populated region have lower family probabilities, as expected. 

\begin{table*}[t]
\centering
\caption{Family Probabilities}
\label{familyness}
\begin{tabular}{cccc}
\hline
\hline
Number & Name & $\Delta v (m s^{-1})$ & Family Probability \\ \hline
55636 & 2002 TX$_{300}$ & $95^{+54}_{-33}$ & $0.999^{+0.001}_{-0.002}$ \\ 
202421 & 2005 UQ$_{513}$* & $139^{+34}_{-26}$ & $0.996^{+0.003}_{-0.007}$ \\ 
145453 & 2005 RR$_{43}$ & $92^{+49}_{-30}$ & $0.994^{+0.003}_{-0.005}$ \\ 
308193 & 2005 CB$_{79}$ & $92^{+47}_{-30}$ & $0.982^{+0.009}_{-0.013}$ \\ 
120178 & 2003 OP$_{32}$ & $143^{+50}_{-32}$ & $0.978^{+0.014}_{-0.043}$ \\ 
19308 & 1996 TO$_{66}$ & $116^{+57}_{-37}$ & $0.978^{+0.011}_{-0.020}$ \\ 
386723 & 2009 YE$_{7}$ & $147^{+57}_{-36}$ & $0.970^{+0.016}_{-0.037}$ \\ 
523645 & 2010 VK$_{201}$* & $143^{+53}_{-35}$ & $0.969^{+0.017}_{-0.031}$ \\ 
\ldots & 2015 AJ$_{281}$* & $114^{+48}_{-30}$ & $0.964^{+0.019}_{-0.033}$ \\ 
315530 & 2008 AP$_{129}$* & $73^{+37}_{-29}$ & $0.954^{+0.026}_{-0.098}$ \\ 
\ldots & 2014 LO$_{28}$* & $82^{+54}_{-32}$ & $0.952^{+0.024}_{-0.038}$ \\ 
416400 & 2003 UZ$_{117}$ & $132^{+58}_{-35}$ & $0.944^{+0.028}_{-0.042}$ \\ 
\ldots & 2014 HZ$_{199}$* & $141^{+38}_{-28}$ & $0.938^{+0.031}_{-0.055}$ \\ 
24835 & 1995 SM$_{55}$ & $174^{+68}_{-45}$ & $0.905^{+0.053}_{-0.124}$ \\ 
\ldots & 2014 QW$_{441}$* & $181^{+59}_{-41}$ & $0.896^{+0.057}_{-0.127}$ \\ \hline
471954 & 2013 RM$_{98}$ & $185^{+76}_{-52}$ & $0.822^{+0.117}_{-0.328}$ \\ 
523776 & 2014 YB$_{50}$ & $188^{+82}_{-50}$ & $0.802^{+0.107}_{-0.199}$ \\ 
523627 & 2008 QB$_{43}$ & $192^{+80}_{-50}$ & $0.769^{+0.123}_{-0.233}$ \\ 
\ldots & 2002 GH$_{32}$ & $209^{+72}_{-43}$ & $0.723^{+0.146}_{-0.256}$ \\ 
471318 & 2011 JF$_{31}$ & $210^{+87}_{-55}$ & $0.310^{+0.473}_{-0.308}$ \\ 
\ldots & 2010 VR$_{11}$ & $252^{+94}_{-64}$ & $0.187^{+0.565}_{-0.187}$ \\ 
\ldots & 2014 NZ$_{65}$ & $361^{+95}_{-61}$ & $0.033^{+0.341}_{-0.033}$ \\ \hline

\end{tabular}
\tablecomments{For the 22 KBOs used to fit the data (Table \ref{tab:KBOs}), we use the posterior distribution from our preferred (quasi-planar PI model) model to self-consistently measure the $\Delta v$ of each KBO relative to the collision center. This results in a posterior distribution for $\Delta v$ which we show here using the median with uncertainties indicating the 16$^{\textrm th}$ and 84$^{\textrm th}$ percentiles. By comparing the likelihood of matching the family model ($\mathcal{L}_{fam}$) to the likelihood of belonging to the background population ($\mathcal{L}_{bg})$, we can also estimate the probability that each KBO is a member of the family as described in the text. This is also performed self-consistently over the posterior distribution. We identify 7 new "dynamically-confirmed" Haumeans with median probabilities greater than 0.9 (indicated by asterisks) in addition to the 8 spectrally-confirmed family members. }

\end{table*}

In addition to finding family probabilities for the KBOs in our dataset, family probabilities can be estimated at any arbitrary point in parameter space. Hence, a 4-dimensional grid of points in $a$-$e$-$i$-$H$ space was created. Each point was evaluated to help build a map of estimated family probabilities for rapid estimation of family probability for future discoveries (though proper elements would still need to be calculated).

\begin{figure}
    \centering
    \includegraphics[width=\textwidth,height=\textheight,keepaspectratio]{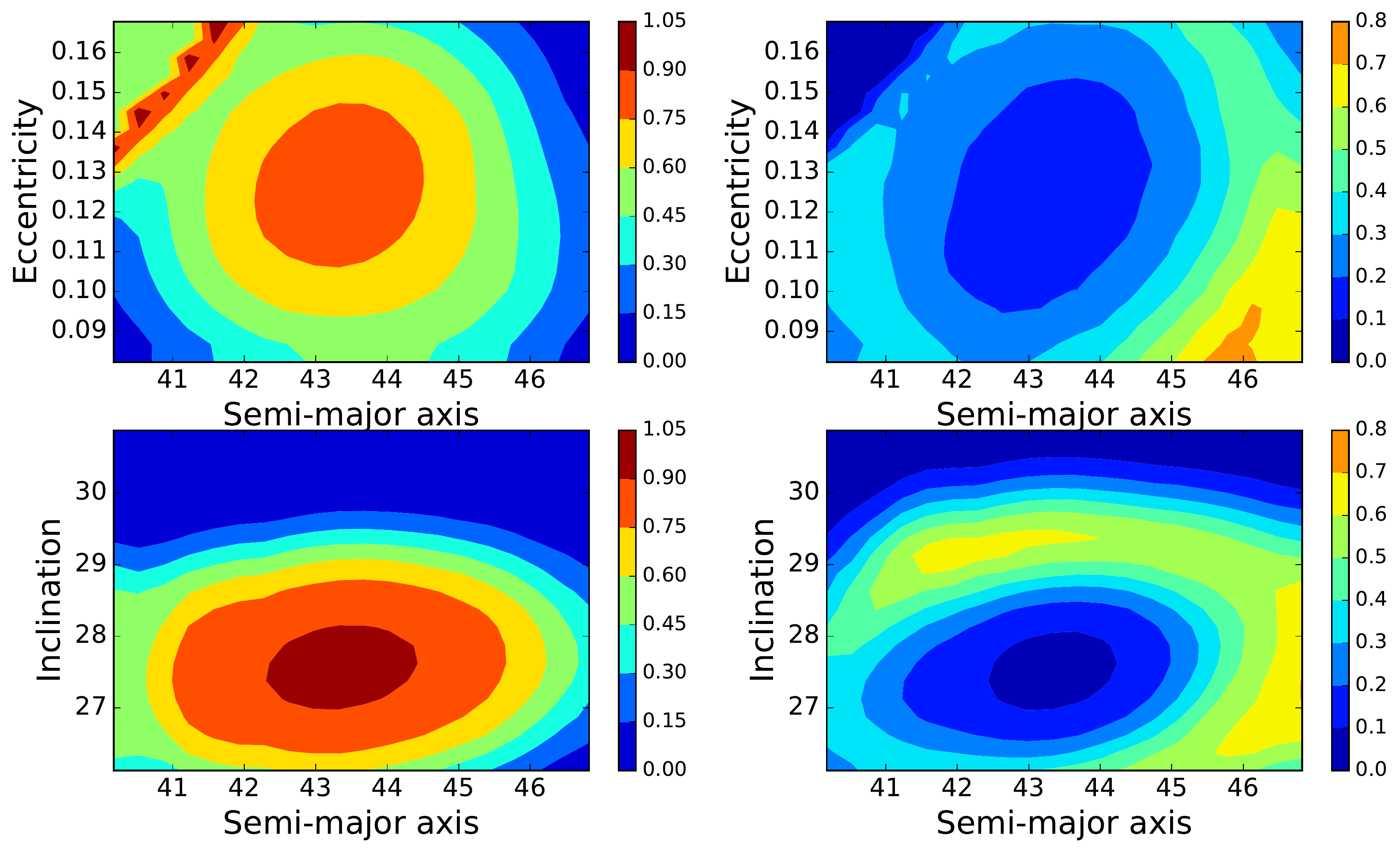}
    \caption{On the left, a contour plot of the median family probabilities is shown. These average over the orbital element not shown (inclination on top and eccentricity on the bottom). On the right, the contour plot showing the interquartile range of the family probabilities. These distributions change as a function of $H$ because of the different absolute magnitude distribution slopes between the family and the background; shown here is the results for $H$=5.74. As can be seen clearly, there is a "core" in which KBO's are likely to be family members. KBOs which lie outside of the core have large uncertainties, as shown in the right panels. Note that these are proper elements rather than osculating elements, the conversion to proper elements can significantly shift the KBO in these plots. In the top two panels, the probabilities in upper left corner are not well-defined due to issues with the background population and the detectability cut. Probabilities in this area are not meaningful and should not be used.}
    \label{fig:supergrid}
\end{figure}

We illustrate how the family probability ($p_{fam}$) appears in $a-e-i$ space (for 5 different $H$ slices) in Figure \ref{fig:supergrid}. At each location in $a-e-i-H$ space, a full posterior distribution of $p_{fam}$ is generated; this Figure shows the median value.\footnote{As in \citet{2007AJ....134.2160R}, one can account for resonant diffusion by allowing a KBO to explore the eccentricity-inclination trade-off allowed when conserving the Tisserand parameter and adopting the highest $p_{fam}$ reached. This would not be a rigorous probability, but it could be used to identify plausible resonant Haumeans.}

We can also show the $a-e-i$ distribution of the uncertainty in $p_{fam}$ as shown in Figure \ref{fig:supergrid}. Here the uncertainty is defined as the interquartile range. This plot emphasizes the locations that are least constrained in parameter space. These correspond to parameter space in which getting new data is most beneficial. 

Both Table \ref{familyness} and Figure \ref{fig:supergrid} show that the family is compact and covers only a small region of parameter space. These also show that the family only dominates at low absolute magnitudes. At higher absolute magnitudes, the family becomes almost indistinguishable from the background. 

\subsection{The Number and Size Distribution of the Haumea Family}
\label{sec:numbersize}
In addition to the model specific parameters providing constraints on collision types, the model parameters shared by all of the family models provide interesting constraints on the absolute size of the Haumea family.

The posterior for the interloper fraction $f_{int}$ gives a range of $\sim$ 0.05-0.4. This indicates that, as expected, there were about 2-8 interlopers in our data, consistent with Table \ref{familyness}. 

With the given interloper fraction ($f_{int}$) and a number of interlopers in the testing range, the total number of Haumea family members can be estimated. The actual number of Haumea family members, in the given $H$ range (3.4-6.0), would be $38_{-14}^{+34}$ family members.

All of the models placed an upper limit of the family's absolute magnitude distribution slope at $\sim$ 0.4. This is significantly lower than the slope for the background population ($\alpha_{bg} = 0.8$). This is consistent with the results of Pike et al. (2018, submitted) who use the Outer Solar System Origins Survey (OSSOS) Simulator to do a much better job at characterizing the detection efficiency. This consistency is an encouraging sign that our simplified detectability cuts were reasonable. This size distribution slope matches the prediction for the graze-and-merge formation hypothesis of \citet{2010ApJ...714.1789L}, but is somewhat more shallow than would be expected for a catastrophic collision \citep{2012ApJ...745...79L}. 

Our interloper fraction $f_{int}$ is integrated over the entire $a-e-i-H$ box. However, since the size distribution slopes for the family and the background population are different, the frequency of interlopers is a function of the absolute magnitude. At low $H$, Haumea family members dominate, due to their shallow $H$ distribution. As $H$ increases, the shallow slope of the Haumea family loses out to the steeper slope of the background population. 

Extrapolating slightly outside our modeled range, we find that the "crossover point" where the number of family members and interlopers are equal is at $H_{cross} \simeq 7$. This was consistent over a wide range of the posterior distribution. At $H \simeq 8.5$, the family will be dominated by the background population 10:1. Around this value of $H$, the size distribution of background objects rolls over to roughly match the $\alpha \simeq 0.4$ for the family (e.g., \citealt{shankman2016ossos}, \citealt{lawler2018ossos}, Pike et al. 2018, submitted). Thus, the 10:1 ratio does not get much worse for $H>8.5$. Note that this is the average ratio over the entire $a-e-i$ box; objects near the center of the family are favored and can have higher probabilities. 

Although our model is conditioned on the non-resonant spectrally-confirmed family members, it identifies family probabilities purely dynamically. Color/spectral properties can then be used to \emph{ad hoc} increase these probabilities. This will be essential for candidate Haumeans with $7 < H < 8.5$. Color/spectral confirmation for such faint objects will typically require the largest telescopes or the James Webb Space Telescope. 

The observed number-size distribution means that most large Haumeans have already been found and that even future deep surveys like the Large Synoptic Survey Telescope (LSST) will result in only modest new numbers of Haumeans, most of which will require significant additional observations to confirm, though we note that LSST is expected to provide phase curve slopes that are diagnostic of spectrally-confirmed Haumeans \citep{rabinowitz2008youthful}. Additional detailed characterization of the family might proceed slowly. For example, the suggestion by \citet{2007AJ....134.2160R} to use the distribution of $\sim$100 12:7 Haumeans to constrain the family formation age will not be observationally feasible in the foreseeable future. 

\section{Discussion}
\label{sec:discussion}

The results do not agree with any of the proposed formation hypotheses at the 95\% (or higher) confidence level. The tightly planar distribution expected from the graze-and-merge formation hypothesis is clearly not seen, even when allowing for a delayed ejection. The velocity boost and anisotropic ejection expected from the satellite formation hypothesis is also clearly disfavored. With these results in hand, we discuss in detail the proposed hypotheses. 

\subsection{Independent Origin Hypothesis}

Returning to the interrelated formation hypotheses illustrated in Figure \ref{fig:hypotheses}, we see that the observed small isotropic $\boldsymbol{\Delta v}$ distribution favors a model where a small object, possibly a former satellite of Haumea, was destroyed in a catastrophic collision \citep{bagatin2016genesis}. A small object is needed to produce the low $\Delta v$'s, a catastrophic collision is needed to produce an essentially isotropic distribution, and the poor match with the satellite collision model indicates that the object was not bound to Haumea at the time. Initially, this seems to be a good match for the Independent Origin Hypothesis of \citet{bagatin2016genesis}.

The center of this observed family is well constrained in both our chosen best fit posterior, produced by the PI model, and in the posteriors of both the other models. The posteriors of all models weakly include the location of the 12:7 resonance in the collision center. However, with the proper to osculating element offset (up to 2 AU), the 12:7 resonance could be included. Thus it remains quite probable that the objects could be ejected from Haumea which later diffuses in the 12:7 resonance.

Decoupling the "Haumea" family from Haumea is considered somewhat probable by \citet{bagatin2016genesis}, but their assessment assumed that Haumea was at its present location at the time of family formation. Since the \emph{present} Haumea and the center of the family are somewhat separated in orbital element space, their model gives a moderate probability that this colocation is a coincidence. However, Haumea is currently in the 12:7 resonance and was likely at lower eccentricity in the past. In particular, the proper Tisserand parameter of Haumea, which is conserved in resonance diffusion, is still consistent with the family center \citep{2007AJ....134.2160R}. Including this constraint would significantly reduce the probability that Haumea is coincidentally near the family, as would be required for the Independent Origin hypothesis. 

A major advantage of the Independent Origin hypothesis is that it can reproduce the tight $\Delta v$ distribution. The tightness of the $\Delta v$ distribution is associated with the fact that all objects with deep water ice signatures are within $\Delta v \lesssim 150$ m s$^{-1}$ of the collision center and that most objects within $\Delta v \lesssim 150$ m s$^{-1}$ of the collision center have strong water ice features (e.g., \citealt{schaller2008detection}, \citealt{carry2012characterisation}). This fact has persisted despite observational surveys of 100+ KBOs that could have detected strong water ice features in a variety of locations (e.g., \citealt{carry2012characterisation}, \citealt{brown2012water}, \citealt{schwamb2018}). Furthermore, the deep water ice spectral features are also associated with "fresh" phase curves \citep{rabinowitz2008youthful} and high to very high albedos (\citealt{elliot2010size}, \citealt{vilenius2018tnos}). Given that the Haumea family is billions of years old and that Haumeans are too small for endogenic geological resurfacing, the most natural explanation for these surface properties is that Haumeans are made from essentially pure water ice. Without any hydrocarbon species to be reddened and darkened by cosmic rays, Haumean surfaces can remain pristine.\footnote{Further modeling implies that the surfaces of Haumeans can remain clean for billions of years despite impacts from red KBOs and accretion of red KBO dust because of the small number of these KBOs as seen in \citealt{2019arXiv190210795S}.} This suggests that the progenitor of the observed family was differentiated or that it originated from a differentiated body. (Others have pointed out that some Haumeans may come from other pieces of Haumea (e.g., the volatile crust) and that these "black sheep" may not be interlopers even if they do not share the same spectral features as the rest of the family. This is discussed in greater detail below.)

We suggest that merely requiring the family progenitor to show water ice features is not a sufficient explanation for these observations and that the family progenitor must have been a rather unique object, though we note that the family is clearly over-represented in the current Kuiper belt because of its very high albedos, implying that pure water ice families may not be so improbable. Still, the uniqueness of the progenitor again reduces the probability that Haumea and the family are independent. 

\citealt{bagatin2016genesis} and \citealt{2012MNRAS.419.2315O} also consider the case that Haumea and the family progenitor are compositionally related in the "quasi-independent origin" hypothesis. However, formation in the same rotational clump in the pre-excitation Kuiper belt is unlikely to lead to the observed family because the delivery to the current location in the Kuiper belt is chaotic and does not preserve the initial location. However, a rotational fission (either from a proto-binary or from the disruption of an already rapidly rotating Haumea by a small KBO impactor) that lead to a single large object could potentially produce a rapidly rotating Haumea and a predominantly icy fragment. \citealt{2012MNRAS.419.2315O} show that this fragment can be immediately ejected or it could be a satellite that, after a gentle ionization, could have similar enough orbital elements to result in an incorrect link between the family and Haumea. The ionization must be quite gentle in order for the collision center to be close to the 12:7 resonance, as discussed above, which makes the pure fission ejection proposed by \citet{2012MNRAS.419.2315O} improbable, as it requires an initial offset velocity of $\sim$400 m $s^{-1}$ (well outside the observed family). Following the formation pathways of asteroid pairs, the suggestion of \citet{2012MNRAS.419.2315O} that tides or other dynamical interactions could lead to the eventual dissolution of the binary would provide a gentle enough ionization. We stress that this formation hypothesis requires that rotational fission produce a single large satellite, which has not been well explored by detailed simulations. 

A crucial drawback to any catastrophic collision of a small body as the origina of the Haumeans is that the velocity distribution of Haumeans is inconsistent with the known mass in these bodies. \citealt{vilenius2018tnos} and Pike et al. 2018, submitted both estimate that the combined mass of the Haumeans is  $\sim$3\% of Haumea's mass (which is $4 \times 10^{21}$ kg, \citealt{2009ApJ...698.1778R}). Since $v_{esc} \propto M^{1/3} \rho^{1/6}$, we can accurately estimate that a single object with this mass would have an escape velocity of $\sim$230 m s$^{-1}$, which would result in a typical ejection velocity for the family of $\gtrsim$500 m s$^{-1}$ \citep{2012ApJ...745...79L}. Former investigations into the Haumea family were satisfied if the typical ejection velocity was equal to the escape velocity. This relied on an incorrect interpretation of \citet{1999Icar..142....5B} which has been cited to say that the typical ejection velocity is 0.7 times the escape velocity; however, this paper is only referencing the ejection velocity of the largest remnant and is not relevant to the family members. \citet{2012ApJ...745...79L} discusses the velocity distribution of ejected family members and finds typical velocities of a few times the escape velocity with a wide range. Since there is such a strong dependence between escape velocity and mass, in order to produce a velocity ejection field similar to the one observed would require an object with mass $\sim$0.1\% of Haumea's mass, much smaller than the known family. Furthermore, the shallow absolute magnitude distribution slope ($\alpha \simeq$ 0.1-0.4) and low size-velocity correlation ($\beta \simeq 0.1$) are not typical outcomes of catastrophic collisions ($\alpha \simeq 0.4$ and $\beta \simeq 0.2$). In short, the distribution of the observed Haumeans cannot be produced in a typical catastrophic collision. 

The dissonance between the ejection distribution inferred by our modeling and our present understanding of collisional outcomes points out the importance of attempting to produce new collisional models guided by these constraints. 

\subsection{Graze-and-Merge Formation Hypothesis}

The graze-and-merge formation hypothesis as described in Section \ref{sec:hypotheses} has many attractive features: it naturally explains Haumea's rapid spin and the presence of a proto-binary (similar to Pluto or Triton) makes the low-velocity collision reasonably probable. 

Is there any way to "save" this model from its failure to match the distribution of $\boldsymbol{\Delta v}$? The tightly planar result would only be consistent with the observations \citep{2010ApJ...714.1789L} 1.1\% of the time; delaying the ejection of family members could change this to 5.3\%, i.e., it is still 2$\sigma$ inconsistent with the observations. As discussed above, we do not believe this difference can be attributed to observational bias. 

To consider whether the addition of new KBOs to the model or unexpected interlopers was to blame, we also explored a PI model fit to only the confirmed members of the Haumea family. While the confidence was lower, this fit also clearly disfavored a planar distribution and was consistent with an isotropic distribution. 

Perhaps there is a graze-and-merge-like formation mechanism that does not produce, in the end, a planar ejection distribution. Additional collision modeling is suggested to match the new constraints discovered herein. However, it is hard to imagine any process that succeeds at shedding angular momentum without creating a strongly preferred ejection direction for the family. Even an extensive period of scattering between objects in the post-merging disk is unlikely to produce the observed typical inclination of $\sim$50$\degr$. 

Another possibility is that some process changed the original planar distribution into the quasi-planar/isotropic distribution seen today. We have already shown that a planar ejection field in the original osculating elements results in the same $a-e-i-H$ shape in proper elements. What about perturbations from Haumea, the large object that is originally on a very similar orbit? We added a massive Haumea at the center of our REBOUND integration and found that it had an almost negligible effect on the orbits and proper elements of the simulated planar Haumeans after 50 MYr. After this point, interactions with Haumea are minimal since all the family members have precessed into orbits with the full range of apses and nodes, so further interactions are not expected.

This simulation did not include Haumean-Haumean interactions or interactions with other KBOs, in the form of gravitational perturbations or collisions (as in \citealt{brovz2018study}). These should be investigated further, but since the interactions with a massive Haumea close to the formation time should be among the strongest effects, but did not make a difference, we consider this somewhat unlikely and leave it for future work.

Weak resonances can cause diffusion in eccentricity and inclination, but preserve the Tisserand parameter. Unfortunately, the $a-T$ distributions of a planar family and an isotropic family are essentially identical, making it difficult to rule out weak resonances. Similarly, it is possible that Haumea formed during the tail of Neptune migration and that some Haumeans were perturbed by Neptune beyond simple resonance diffusion (e.g., transport in semi-major axis during resonance capture). We do note that the proper semi-major axis distribution of high-probability Haumeans has gaps from 41.8-42.8 and 43.4-44.2; though these are presumably due to the 5:3 resonance at 42.27 and the 7:4 resonance at 43.67, respectively, the gaps are much larger than can be explained simply with the resonances at their present locations. 

We have not modeled these scenarios, but matching the distribution of Haumeans poses a difficult challenge: an originally planar Haumea family distribution needs to be "shaken" enough to appear nearly isotropic today, but not so much to change the tight $\Delta v$ distribution. Finding a probable model on this knife-edge of perturbation seems difficult. Proving that it is a better fit to the family than an isotropic distribution will require detailed modeling. If such a model were successful, it could have the additional advantage of revealing useful information about the timescale for Haumea family formation and/or the constraints on the migration of Neptune. 

The creation of a secondary family (e.g., the catastrophic destruction of a family member) in the presence of a planar distribution also seems unlikely since the ability of our model to discard interlopers suggests that any original truly planar distribution must be a fraction (perhaps half at most) of the present-day family. This scenario would also fall subject to the same issues with the catastrophic disruption of this "independent" object discussed above.

Despite these challenges, if a natural explanation can be found that can connect the graze-and-merge formation hypothesis with the quasi-planar distribution of Haumeans, the rest of the story could fall neatly into place. 

\subsection{Satellite Collision Hypothesis}

Our modeling disfavors the satellite collision hypothesis because we have shown that satellite collisions can be recovered in tests, but application to the actual data resulted in a posterior distribution that attempted to match an isotropic distribution. The significant $\Delta BIC$ between the satellite collision model and the PI model is also good rationale for disfavoring a satellite collision. Furthermore, the "boost" from the satellite's orbital velocity is likely to shift the collision center significantly, making it less likely to connect to the 12:7 resonance that Haumea currently occupies. 

The satellite collision hypothesis has other challenges as well. It does not naturally connect Haumea's rapid spin with Haumea's family, leading to an overall improbable story (e.g., \citealt{bagatin2016genesis}). Furthermore, SPH modeling shows that the initial proposed step (the formation of a large satellite and rapid spin) is not plausible \citep{2010ApJ...714.1789L,2012MNRAS.419.2315O}.\footnote{One area for possible investigation with the satellite hypothesis is to consider whether SPH simulations of a graze-and-merge-like impact could shed angular momentum by ejecting a single large object instead of many small fragments.} Said another way, Haumea's spin period is a clear outlier for all large bodies in the solar system, which suggests a unique process is needed beyond collisions. Indeed, while \citet{cuk2013dynamics} provide some justification for why the tidal expansion of the large ur-satellite could leave enough spin angular momentum in Haumea; this implies that Haumea's spin used to be even closer to breakup in the past. 

As with the independent catastrophic collision model, the satellite collision hypothesis will also struggle to match the number-size-velocity distribution of the observed objects since the total family mass is too large to produce such a low velocity and may not be consistent with the observed $\alpha$ and $\beta$. 

\citet{cuk2013dynamics} favor the satellite collision hypothesis since it may provide better initial conditions for the large semi-major axes of Haumea's moons. (This is not consistent with the small ejection velocities we observed earlier as the progenitor to these moons would have had Haumea-centric velocity of $\sim$100 m s$^{-1}$.) Starting with a distant disk avoids the problematic tidal properties that would be required to attain the observed semi-major axes \citep{quillen2016tidal}. Even so, there is enough uncertainty in the dynamical understanding of tides that this rationale for the satellite collision may not be compelling. Whether such a distant disk would lead to satellites like those observed (which are quite well separated) is also not clear. Furthermore, even the requirement of an initially distant disk does not immediately imply a satellite collision; the properties of the proto-satellite disk after a graze-and-merge collision are poorly understood and might be sufficient to form distant satellites. 

For these reasons, we believe that the satellite collision hypothesis is unlikely. Further modeling would need to address the key issues raised here. 
\subsection{Other Formation Hypotheses}
\label{sec:otherhypotheses}
Our rejection of the predominant hypotheses has led us to consider other possibilities. 

\subsubsection{Cratering Collision}
With only $\sim$3\% of Haumea's mass ejected in the collision, another hypothesis is that Haumea's family is the result of a "cratering" collision. In this sub-catastrophic collision, the formation of a large crater on Haumea results in the ejection of a small amount of mass on potentially low velocity orbits. This smaller collision would be relatively probable to occur in the age of the solar system. However, additional SPH modeling would be needed to see if it could produce ejecta at just above the escape velocity (so that they have small $\Delta v$), two relatively large moons on nearly circular coplanar orbits, and/or the near break-up spin. 

We investigated the $\boldsymbol{\Delta v}$ model that would result from such a collision. For simplicity, we consider the debris that would be ejected from a flat plane, ignoring Haumea's curvature and tri-axial shape as is likely reasonable for such a small collision. Ejecta from this impact would leave in a cone around the normal to this plane. We considered model families that included all the general free parameters in the simulations above and which ejected family members in a cone with a variable direction, variable opening angle, and variable scatter around this angle. The result is a highly directional ejection field. In the spherical equal area projection of the ejection vectors, this collision creates a circle-like shape with size proportional to the opening angle (assumed to be relatively small as is known for craters). This is a poor match visually to the observations and attempts to fit this model to the data led to poor likelihoods. In addition, the highest likelihood regions had cone angle openings of 180$\degr$, at which point the model becomes degenerate with the PI model. While we did not perform the extensive testing on this model that we did for our other models, the distribution of Haumeans does not seem to support this model.

Given the observational and theoretical challenges of a cratering collision, we conclude that it is not a viable hypothesis at this time. 

\subsubsection{Tip of the Iceberg Hypothesis}

A large isotropic collision is a good match to the ejection distribution of Haumeans, but an awful match to the observed tight ejection velocities. What if the observed tight spectrally-confirmed Haumeans are merely the "tip of the iceberg" of a much larger family? When larger families are invoked, we can relax the limitations against a catastrophic collision.

Consider a mantle-stripping collision of a proto-Haumea with radius 1300-1500 km. In a simple zeroth-order model, suppose that each piece of the proto-Haumea is given an equal initial energy, such that objects lower in the body have a greater potential well to escape from and thus have smaller $\Delta v$'s.

We have not used geophysical constraints to construct this proto-Haumea and instead attempt to place empirical constraints. A possible internal structure is as follows: a rocky core, surrounded by a solid water-ice mantle, then a liquid water ocean, and a dusty crust. This collision ejects the dusty crust at high velocity ($\sim$500-1000m s$^{-1}$). The liquid water ocean in the process of the collision is vaporized and then recrystallizes into small, undetectable ice crystals that are then removed by radiation pressure as in \citet{chau2018forming}. Finally, part of the water ice mantle is ejected at the low velocities of spectrally-confirmed Haumeans.  Beneath this, there is insufficient energy for any more mass to ultimately escape.

\begin{figure}
    \centering
    \includegraphics[width=0.5\textwidth,height=0.5\textheight,keepaspectratio]{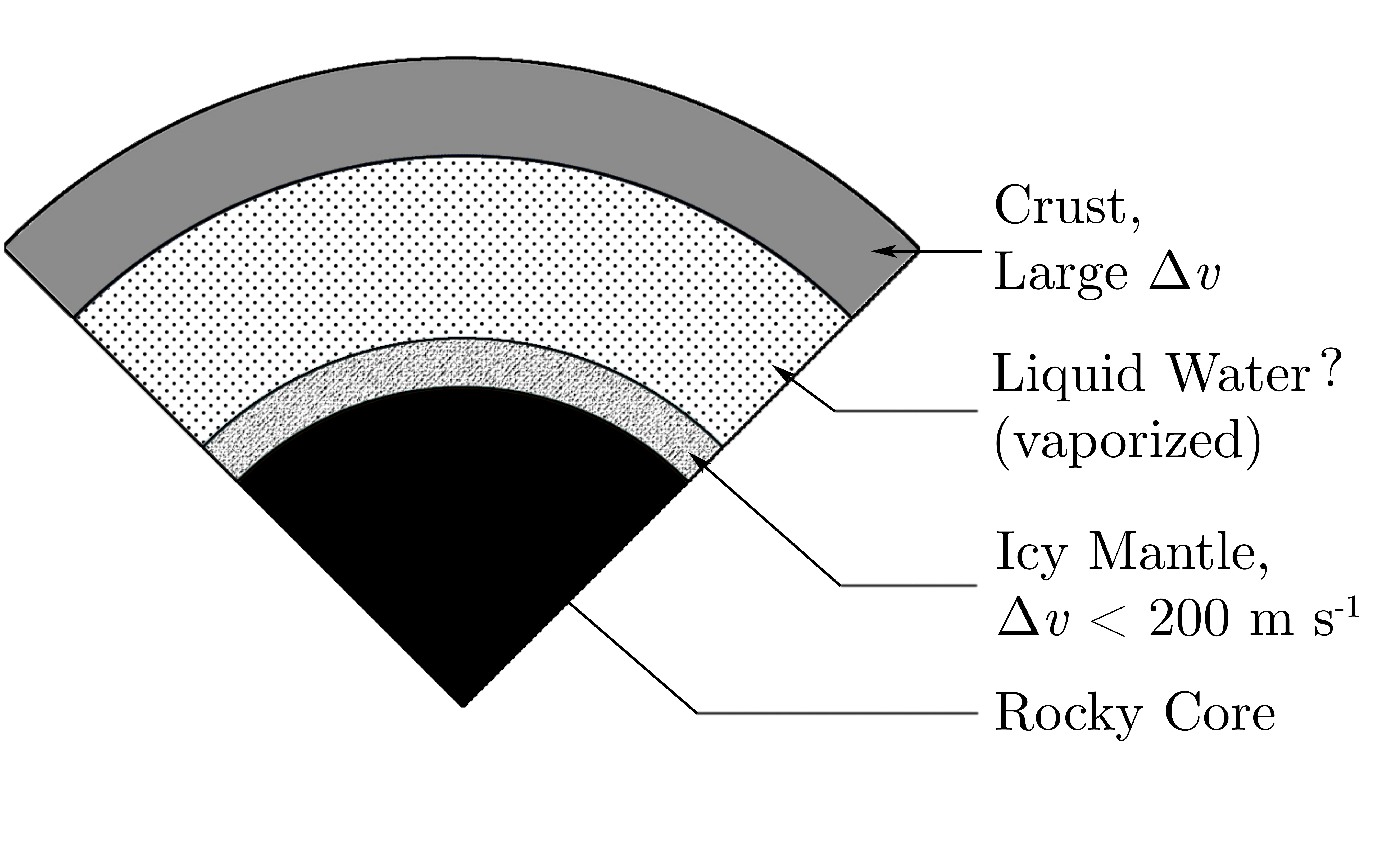}
    \caption{One possible realization of the interior of the proto-Haumea in the Tip of the Iceberg hypothesis. In this model, the proto-Haumea consists of four different layers, a rocky core, an icy mantle, a possible liquid water layer, and an outer crust. A catastrophic collision ejects the crust at a large $\Delta v$, rendering it so widespread as to make it undetectable. The liquid water layer (if any) is vaporized to reform small ice crystals that cannot be detected. Finally the last ejected layer, an icy mantle, is ejected at a low $\Delta v$, creating the family seen today. All layers under this eventually reform into the Haumea seen today. This new hypothesis may be consistent with some observations, but has several serious issues that require further investigation.}
    \label{fig:toti}
\end{figure}

When the proto-Haumea is $\sim$1300-1500km in radius, the mass ejected within 200m s$^{-1}$ is $\sim$ 10\% of Haumea's mass, about equal to current family mass estimates. Fragments below this layer have a mass equal to Haumea's present mass; they do not escape and eventually reaccumulate to form Haumea. The fraction of ocean to "crust" mass is unconstrained. In this oversimplified model, some of the traits of the Haumea family can be naturally explained: the nearly pure water-ice Haumeans with a low ejection velocity and a largest remnant with a high density. We have also confirmed that a single self-consistent model is possible, illustrated in Figure \ref{fig:toti}. In addition, the results of \citet{ortiz2017size} could be used to provide further constraints on a possible geophysical model for this hypothesis. We will leave this to others with more expertise in this area.

The pieces of the dusty crust that are ejected at high velocity are scattered across the entire Kuiper Belt (and beyond), $a \simeq$ 30-100AU, $e \simeq$ 0.0-0.6, and $i \simeq$ 15-40$\degr$ making them near impossible to find as they become incredibly widespread. Because all of these fragments need to intersect the collision distance $r$, the resulting distribution in $a-e$ phase space is similar to $a(1+e) = r$ for $a<r$ and $a(1-e) = r$ for $a>r$. Additionally, if the dusty crust of the proto-Haumea is similar to the spectral features of other KBOs, the fragments of the crust would have no unique spectral features to identify them as family members. Such objects likely have albedos $\sim$10 times smaller than the known Haumeans. 

Can we rule out a dark distant cloud of Haumea fragments? We searched for niches of $a-e-i-H$ parameter space where such a collision would produce large numbers of objects that are not seen, focusing on Haumea's high inclination. No such niches were identified; for example, there are significant numbers of $i \simeq 30-40\degr$ KBOs that cannot be \emph{a priori} discarded. This is true even if the mass of dark fragments is much higher than the mass of the water ice fragments. We note that if the mass of the dark fragments is several times the mass of the known Haumea family, then it should contain some objects that are hundreds of kilometers in radius which would probably be known, even if they had lower albedos. 

Modeling will be required to see if a geophysically-plausible proto-Haumea can actually match the detailed observations: ejection of a nearly-pure water ice chunks in a nearly isotropic low $\Delta v$ distribution with the right number-size-velocity distribution; possible vaporization/recrystallization of a liquid water ocean; relatively distant ejection of crust pieces; final spin near break up; and a (possibly-distant) disk for forming Haumea's moons. Some concerns with this model are immediately apparent.
\begin{itemize}
    \item As noted earlier, catastrophic collisions do not result in near-break up spins.
    \item Except for a very narrow range of parameters, it is not expected that a liquid ocean layer would overlay a solid water ice layer (since the latter would be at higher temperatures but only marginally higher pressure).
    \item The equipartition of energy throughout the body as a function of depth would have to be obeyed very well to keep the water ice fragments confined to such a narrow range of $\Delta v \ll v_{esc}$. 
    \item The shallow size-distribution slope of water ice family members is not expected for a catastrophic collision.
    \item There is tension between the idea that the collision is "catastrophic" and that the collision only excavates down to a certain depth. The latter is required to keep the water ice fragments at a low mass and $\Delta v$. A sub-catastrophic mantle-stripping collision may not produce the desired near-isotropic $\boldsymbol{\Delta v}$ distribution.
    \item \citet{2008AJ....136.1079L} are able to use the collision of two scattered disk objects to make the Haumea collision somewhat probable. This updated model requires a much larger proto-Haumea and larger impactor, decreasing the collision probability significantly. Note that the proto-Haumea in this case would be Triton sized or larger. 
\end{itemize}

Additional modeling is required to determine whether such barriers could be overcome. Our sense is that a self-consistent model would be elusive, but given the failure of other models, more complex formation hypotheses need to be explored. 

\section{Conclusions}
\label{sec:conclusions}

We apply Bayesian parameter inference to ascertain the properties of the Haumea family under different assumed models: the PI model, the DEPI model, and the satellite collision model. These models have many flexible parameters and allow for a variable size background/interloper population for robustness. Tests on simulated families showed that these models can recover parameters and can be used to accurately discriminate between models when they do not overlap. 

Efficient exploration of these models allow for many millions of possible family parameters to be explored in order to search for the posterior distributions of these parameters when attempting to match the $a-e-i-H$ distribution (equivalent to the $\boldsymbol{\Delta v}$ distribution) of observed KBOs in a range that includes known Haumeans. 
We find that the model can reject the planar ejection field that is characteristic of the graze-and-merge formation hypothesis (and similar hypotheses). Our preferred fit has a typical inclination width of $\sim$50$\degr$, e.g., it is quasi-planar and consistent with isotropic. This makes it difficult to connect Haumea's rapid rotation with its family. Including a delayed-ejection aspect did not "rescue" this model. Though many aspects of the graze-and-merge formation hypothesis are very appealing, without some additional mechanism to subtly disperse the Haumeans, it cannot yet be considered as a viable self-consistent hypothesis. 

We find that the preferred solution for the satellite collision model is essentially the same as the near-isotropic solution above. The model rejects satellites with properties expected for Haumea's ur-satellite. Additional issues with the satellite collision hypothesis suggest that it is overall unlikely. In particular, the known mass of the Haumeans, the observed velocity distribution, and the overall number-size-velocity distribution are not consistent with catastrophic collision models. This is a serious challenge for all the proposed hypotheses which invoke destruction of a small body, whether a satellite, ionized satellite, or independent object.

Adopting the planar-isotropic model as an empirical distribution of family members, we can calculate posterior distributions for the family probability for any (proper) $a,e,i,H$. This is used to identify the probabilities for a variety of new family members and we dynamically confirm 2005 UQ$_{513}$, 2010 VK$_{201}$, 2015 AJ$_{281}$, 2008 AP$_{129}$, 2014 LO$_{28}$, 2014 HZ$_{199}$, and 2014 QW$_{441}$ as family members. We also identify a parameter space where spectral confirmation of candidate Haumeans would be most valuable (Figure \ref{fig:supergrid}). 

Having strongly disfavored the existing formation models (Figure \ref{fig:hypotheses}), we consider a cratering collision model which is also a poor fit to the data. We also propose a new hypothesis, "Tip of the Iceberg", in which the observed spectrally-confirmed family members are actually a small part of a larger (darker, more distant) family. The other component of this family is difficult to detect because of the removal of a possible liquid water ocean layer and/or the extreme orbital element range of ejected fragments. As there are many apparent drawbacks to this model, it would need to be explored with more geophysical and collisional modeling to be considered seriously. 

With the addition of new Haumeans and new modeling, the mystery of Haumea's formation deepens. We call on other researchers to investigate various aspects of this problem in more detail, especially collision simulations. Understanding icy collisions in the Kuiper belt -- for which Haumea currently provides the tightest constraints -- is a central aspect of understanding core accretion in the outer solar system. 

\acknowledgements
We gratefully acknowledge Steven Maggard for working diligently on orbital integrations of KBOs to determine the proper elements on which this work relied. We thank Rosemary Pike, Mike Brown, David Nesvorny, Nate Benfell, Brooke Crookston, Corey Hawkins, and others for useful discussions. We acknowledge funding support from an internal Brigham Young University Mentoring Environment Grant and funding from the Division for Planetary Sciences (DPS) Hartmann Travel Grant so that this work could be presented at the 2018 DPS meeting. Lucidchart was used to prepare Figure \ref{fig:hypotheses}. 

\bibliographystyle{aasjournal}

\end{document}